\documentclass[preprint,12pt,authoryear]{elsarticle}

\usepackage{amssymb}
\usepackage{graphicx}%
\usepackage{multirow}%
\usepackage{amsmath,amssymb,amsfonts}%
\usepackage{amsthm, url}%
\usepackage{mathrsfs}%
\usepackage[title]{appendix}%
\usepackage{xcolor}%
\usepackage{textcomp}%
\usepackage{manyfoot}%
\usepackage{booktabs}%
\usepackage{algorithm}%
\usepackage{algorithmicx}%
\usepackage{algpseudocode}%
\usepackage{listings}%
\usepackage{lscape}
\usepackage{setspace}
\usepackage{lipsum}
\journal{Spatial Statistics}

\begin{document}

\setlength{\tabcolsep}{2pt}
\begin{frontmatter}



\newtheorem{theorem}{Theorem}
\newtheorem{assumption}{Assumption}
\newtheorem{proposition}{Proposition}
\newtheorem{corollary}{Corollary}
\newtheorem{lemma}{Lemma}

\title{SAR models with specific spatial coefficients and heteroskedastic innovations}


\author[label1]{N.A. Cruz\corref{cor1}}
\affiliation[label1]{organization={Profesor Visitante, Universitat de les Illes Balears, Departament de Matemàtiques i Informàtica, Phone: +34 637 54 6888, Palma de Mallorca, España},
ead={nelson-alirio.cruz@uib.es}, ead={https://orcid.org/0000-0002-7370-5111}}

\author[label2]{D.A. Romero}
\affiliation[label2]{organization={Departamento de Estadística, Facultad de Ciencias,  Universidad Nacional de Colombia, Bogota, Colombia},
ead={dfdelgadoc@unal.edu.co}}

\author[label3]{O.O. Melo}
\affiliation[label3]{organization={Departamento de Estadística, Facultad de Ciencias,  Universidad Nacional de Colombia, Bogota, Colombia},
ead={oomelo@unal.edu.co}, ead={https://orcid.org/0000-0002-0296-4511}}

\begin{abstract}
This paper presents an innovative extension of spatial autoregressive (SAR) models, introducing spatial coefficients specific to each spatial region that evolve over time. The proposed estimation methodology covers both homoscedastic and heteroscedastic data, ensuring consistency and efficiency in the estimators of the parameters $\pmb{\rho}$ and $\pmb{\beta}$. The model is based on a robust theoretical framework, supported by the analysis of the asymptotic properties of the estimators, which reinforces its practical implementation. To facilitate its use, an algorithm has been developed in the R software, making it a standard tool for the analysis of complex spatial data.
The proposed model proves to be more effective than other similar techniques, especially when modeling data with normal spatial structures and non-normal distributions, even when the residuals are not homoscedastic.
Finally, the application of the model to homicide rates in the United States highlights its advantages in both statistical and social analysis, positioning it as a key tool for the analysis of spatial data in various disciplines.
\end{abstract}


\end{frontmatter}

\section{Introduction}\label{sec1}

Spatial autoregressive (SAR) models, proposed by \citet{cliff1973spatial}, have attracted the attention of researchers due to their ability to model the spatial dependence that arises when values in a region or location are related to neighboring observations, reflecting this dependence through an autoregressive parameter associated with the dependent variable. This approach solves the overspecification problem that arises when trying to specify unique coefficients for each spatial region in a cross-sectional dataset, a case where the available observations are not sufficient to carry out the estimation. The generic structure of a SAR model, popularized by \citet{anselin2013spatial} is given by:
\begin{equation}
y_{i} = \rho \sum_{j=1}^N w_{ij}y_{j} + \sum_{k=1}^K x_{ik}\beta_{k} + \epsilon_{i}
\end{equation}
where $y_{i}$ is the dependent variable of spatial region $i$, $w_{ij}$ is the spatial weight relating unit $i$ to unit $j$, and $x_{ik}$ are the explanatory variables associated with unit $i$. The parameter $\rho$ reflects the magnitude of the spatial dependence, while $\beta_{k}$ are the coefficients of the explanatory variables. The main objective of SAR models is to find the parameters corrected for the spatial dependence existing between the analyzed units, which is not completely explained by the variables in $\sum_{k=1}^K x_{ik}\beta_{k}$, and to measure the magnitude of such dependence.

The recent rise of cross-sectional databases observed in successive periods of time (known as panel data) has attracted attention in the field of spatial models, since they allow greater variability and degrees of freedom across the temporal dimension. This has led to the creation of extensions to the SAR models for panel data, such as the SAR model for panel data (SARP), which adopts the structure described below:
\begin{equation}
\mathbf{Y}= \rho(\mathbf{I}_{T}\otimes\mathbf{W})\mathbf{Y} + \mathbf{X}\pmb{\beta} + \pmb{\epsilon}
\end{equation}
where $\mathbf{Y}$ is a random vector of dimension $nT \times 1$ containing the response variables, $\mathbf{W}$ is the spatial weight matrix of dimension $n\times n$, and $\mathbf{X}$ is a matrix of dimensions $nT \times K$ containing the $K$ explanatory variables. The parameter $\rho$ still represents spatial dependence and $\pmb{\beta}$ is the vector of marginal effects of the covariates.

Over the years, the literature on spatial models for panel data has experienced significant growth, with new proposals that complement the base specifications of these models. Important examples of these contributions include the work of \citet{baltagi2003testing}, which considered tests to assess spatial dependence in the error component of panel data, and that of \citet{kapoor2007panel}, which introduced the flexibility of specifying random effects in spatial regions. In addition, \citet{lee2010estimation} and \citet{mutl2008spatial} proposed estimation methods for spatial models with both fixed and random effects. 

\citet{regis2022random} presented a spatio-temporal methodology with random effects, \citet{porter2024objective} constructed a spatio-temporal Bayesian model with common autoregressive parameters for the entire space. \cite{li2024higher} introduced a higher-order spatial autoregressive varying coefficient model where regression coefficients are allowed to smoothly change over space; however, the correlation parameters are associated with different spatial weight matrices and not with each spatial region.

Despite these advances, there are no proposals in the literature that allow modeling the specific spatial autoregressive effects of each region on the responses. This theoretical and methodological gap is what motivates the present work, which seeks to propose a new functional form that integrates the specific spatial effects of each unit in the model and develops an adequate estimation process. The estimators of the spatial model with specific coefficients are obtained under the assumption of homoscedasticity. This methodology is extended to a robust one when there is heteroscedasticity in the perturbations of the model. The theoretical asymptotic results for the estimators are obtained, as well as the computational form of all the results. The simulation demonstrates the advantages and problems of the new method in homoscedastic and normal scenarios and non-normal heteroscedastic scenarios. The example of the application of crime patterns in the United States shows the goodness of the proposal.

\section{SAR models with specific spatial coefficients}

Let ${Y}_{it}$ be a random variable associated to $i$-th spatial region in $t$-th time. A spatial autoregressive model with specific spatial coefficient can be specified as:
\begin{align}
    {y}_{it}&=\sum_{j=1}^n\rho_{j} w_{ij}y_{jt} + \sum_{k=1}^K \beta_{k} x_{itk}  + \epsilon_{it}\label{PSAR}
\end{align}
where $i=1, \ldots, n$, $t=1, \ldots, T$, $\rho_i\in (-1,1)$ is a spatial autoregressive parameter associate to spatial region $i$, $\mathbf{W}$ is the row-standardized spatial weight matrix, i.e., $\mathbf{W}\mathbf{1}_n = \mathbf{1}_n $ and $w_{ij}\geq 0$ \citep{anselin1988spatial}, $\pmb{\beta}=(\beta_1, \ldots, \beta_K)^\top$ is a vector of size $K$ with the effects of the $K$ explanatory variables constituting the vector $\mathbf{x}_{it}=({x}_{it1}, \ldots, {x}_{itK})$.
The matrix form of the model proposed in Equation \eqref{PSAR} is:
\begin{align}
\mathbf{Y} &=(\mathbf{I}_T\otimes \mathbf{W})(\mathbf{I}_T\otimes \mathbf{P})\mathbf{Y} +\mathbf{X}\pmb{\beta}+\pmb{\epsilon}\nonumber\\
\mathbf{Y} &=(\mathbf{I}_T\otimes \mathbf{WP})\mathbf{Y} +\mathbf{X}\pmb{\beta}+\pmb{\epsilon}\nonumber\\
(\mathbf{I}_{nT}-(\mathbf{I}_T\otimes \mathbf{WP}))\mathbf{Y} &=\mathbf{X}\pmb{\beta}+\pmb{\epsilon}\nonumber\\
\mathbf{AY} &=\mathbf{X}\pmb{\beta}+\pmb{\epsilon}\label{MatrixPSAR}
\end{align}
with
\begin{equation}\nonumber
      \pmb{Y}=\begin{bmatrix}
        \mathbf{Y}_1\\
        \vdots\\
        \mathbf{Y}_t\\
        \vdots\\
        \mathbf{Y}_T
    \end{bmatrix}, \;
    \pmb{Y}_t=\begin{bmatrix}
        {Y}_{1t}\\
        \vdots\\
        {Y}_{it}\\
        \vdots\\
        {Y}_{nt}
    \end{bmatrix}, \;
\pmb{\epsilon}= \begin{bmatrix} \pmb{\epsilon}_{1} \\ \pmb{\epsilon}_{2} \\ \vdots \\ \pmb{\epsilon}_{T} \end{bmatrix}, \text{ and }\; \mathbf{P}=\begin{bmatrix} \rho_{1} & 0 & \cdots & 0 \\
0 & \rho_{2} & \cdots& 0 \\
\vdots & \vdots & \ddots & \vdots \\
0 & \cdots & \cdots & \rho_{n}
\end{bmatrix}\nonumber\\
\end{equation}
and $\otimes$ indicates the Kronecker product \citep{harville1998matrix}. Assuming that $\forall i=1, \ldots , n, \; \rho_i\in (-1,1)$, the matrix $\mathbf{A}=\mathbf{I}_{nT}-(\mathbf{I}_T\otimes \mathbf{WP})$ is not singular.
\subsection{Homoscesdatic SAR models with specific spatial coefficients}\label{hoSPSAR}
Based on the  normal distribution of the error term  $\epsilon_{it}\sim N(0, \sigma^2)$ and using Equation \eqref{MatrixPSAR}, the log-likelihood function for the observations vector $\mathbf{Y}$ will be:
$$\ell(\pmb{\beta}, \pmb{\rho}, \sigma^2)=-\frac{nT}{2}\ln(2\pi\sigma^2)+\ln|\mathbf{A}|-\frac{1}{2\sigma^2}(\mathbf{AY}-\mathbf{X}\pmb{\beta})^\top
(\mathbf{AY}-\mathbf{X}\pmb{\beta})$$  
The partial derivatives of $\ell$ with respect to the different parameters of the model are given by:
\begin{align}
\frac{\partial{\ell}}{\partial\boldsymbol{\beta}}&=\frac{1}{\sigma}(\mathbf{AY}-\mathbf{X}\pmb{\beta})^{\top}\mathbf{X}\nonumber\\
\frac{\partial{\ell}}{\partial{\sigma^2}}&=-\frac{nT}{2\sigma^2}-\frac{1}{2\sigma^4}(\mathbf{AY}-\mathbf{X}\pmb{\beta})^\top
(\mathbf{AY}-\mathbf{X}\pmb{\beta})\nonumber\\
\frac{\partial{\ell}}{\partial{\rho_i}}&=-T\left[(\mathbf{I}_n\otimes \mathbf{WP})^{-1}\mathbf{W}\right]_{ii}+\frac{1}{\sigma^2}\left[(\mathbf{AY}-\mathbf{X}\pmb{\beta})^{\top}\left(\mathbf{I}_T\otimes \mathbf{W}\mathbf{e}_i\mathbf{e}_i^\top\right)\right]_{ii}\label{gradrho}
\end{align}
where $\mathbf{e}_i$ is the $i$ basis vector of $\mathbb{R}^n$. The deduction of \eqref{gradrho} is shown in \ref{apenA}. Equating to zero these equations and solving, it is obtained that:
\begin{align}
    \hat{\pmb{\beta}}\vert \pmb{\rho}&=(\mathbf{X}^{\top}\mathbf{X})^{-1}\mathbf{X}^{\top}\mathbf{AY}=(\mathbf{X}^{\top}\mathbf{X})^{-1}\mathbf{X}^{\top}[\mathbf{I}_{nT}-(\mathbf{I}_{T}\otimes\mathbf{WP})]\mathbf{Y}\label{betarho}\\
    \hat{\sigma}^2\vert\hat{\pmb{\beta}}, \pmb{\rho}&=\frac{1}{nT}(\mathbf{AY}-\mathbf{X}\hat{\pmb{\beta}})^\top
(\mathbf{AY}-\mathbf{X}\hat{\pmb{\beta}})\label{sigmabetarho}
\end{align}
There is no closed solution to maximize the likelihood function in terms of $\pmb{\rho}$. 
 However, it is clear that:
\begin{align}
    \mathbf{AY}-\mathbf{X}\hat{\pmb{\beta}} &=\mathbf{AY}-\mathbf{X}(\mathbf{X}^{\top}\mathbf{X})^{-1}\mathbf{X}^{\top}\mathbf{AY}\nonumber\\
    &=(\mathbf{I}_{nT}-\mathbf{X}(\mathbf{X}^{\top}\mathbf{X})^{-1}\mathbf{X}^{\top})\mathbf{AY}\nonumber\\
    &=(\mathbf{I}_{nT}-\mathbf{H})\mathbf{AY}
\end{align}
where $\mathbf{H}=\mathbf{X}(\mathbf{X}^{\top}\mathbf{X})^{-1}\mathbf{X}^{\top}$ is the projection matrix in linear models \citep{harville1998matrix}. Therefore, if the estimators obtained in equations \eqref{betarho} and \eqref{sigmabetarho} are replaced in the log-likelihood function, it is obtained that:
\begin{align}
    \ell(\hat{\pmb{\beta}}, \pmb{\rho}, \hat{\sigma}^2)&=-\frac{nT}{2}\ln\left[\frac{1}{nT}(\mathbf{AY}-\mathbf{X}\hat{\pmb{\beta}})^\top
(\mathbf{AY}-\mathbf{X}\hat{\pmb{\beta}})\right]+\ln|\mathbf{A}|-\frac{nT}{2}(1+\ln{(2\pi)})\nonumber\\
&=-\frac{nT}{2}\ln\left(((\mathbf{I}_{nT}-\mathbf{H})\mathbf{AY})^\top
(\mathbf{I}_{nT}-\mathbf{H})\mathbf{AY})\right)+\ln|\mathbf{A}| + C\label{newrho}
\end{align}
where $C=-\frac{nT}{2}(1+\ln{(2\pi)})$ is a constant independent of $\pmb{\rho}$. Therefore, if the gradient and Hessian of \eqref{newrho} is calculated, respectively, as:
\begin{align}
    &\frac{\partial\ell(\hat{\pmb{\beta}}, \pmb{\rho}, \hat{\sigma}^2)}{ \partial\rho_i}=\frac{nT}{2}[\mathbf{Y}^\top\mathbf{A}^\top(\mathbf{I}_{nT}-\mathbf{H})\mathbf{AY}]^{-1}\Big[\operatorname{tr}((\mathbf{I}_{nT}-\mathbf{H})(\mathbf{I}_T\otimes \mathbf{W}\mathbf{e}_i\mathbf{e}_i^\top\mathbf{YY}^\top\mathbf{A}^\top))+\nonumber\\
    &\operatorname{tr}(\mathbf{I}_{nT}-\mathbf{H})\mathbf{AYY}^\top(\mathbf{I}_T\otimes \mathbf{e}_i\mathbf{e}_i^\top\mathbf{W}^\top)\Big] -T((\mathbf{I}_n-\mathbf{WP})^{-1}\mathbf{W})_{ii}\label{drho}
\end{align}
and
\begin{align}  
    & \frac{\partial^2\ell(\hat{\pmb{\beta}}, \pmb{\rho}, \hat{\sigma}^2)}{ \partial\rho_i\partial\rho_j}=-\frac{nT}{2}[\mathbf{Y}^\top\mathbf{A}^\top(\mathbf{I}_{nT}-\mathbf{H})\mathbf{AY}]^{-2}\Big[\operatorname{tr}((\mathbf{I}_{nT}-\mathbf{H})(\mathbf{I}_T\otimes \mathbf{W}\mathbf{e}_i\mathbf{e}_i^\top\mathbf{YY}^\top\mathbf{A}^\top))+\nonumber\\
    &\operatorname{tr}(\mathbf{I}_{nT}-\mathbf{H})\mathbf{AYY}^\top(\mathbf{I}_T\otimes \mathbf{e}_i\mathbf{e}_i^\top\mathbf{W}^\top)\Big]\Big[\operatorname{tr}((\mathbf{I}_{nT}-\mathbf{H})(\mathbf{I}_T\otimes \mathbf{W}\mathbf{e}_j\mathbf{e}_j^\top\mathbf{YY}^\top\mathbf{A}^\top))+\nonumber\\
    &\operatorname{tr}(\mathbf{I}_{nT}-\mathbf{H})\mathbf{AYY}^\top(\mathbf{I}_T\otimes \mathbf{e}_j\mathbf{e}_j^\top\mathbf{W}^\top)\Big]-\nonumber\\
    &\frac{nT}{2}[\mathbf{Y}^\top\mathbf{A}^\top(\mathbf{I}_{nT}-\mathbf{H})\mathbf{AY}]^{-1}\Big[\operatorname{tr}((\mathbf{I}_{nT}-\mathbf{H})(\mathbf{I}_T\otimes \mathbf{W}\mathbf{e}_i\mathbf{e}_i^\top\mathbf{YY}^\top(\mathbf{I}_T\otimes \mathbf{e}_j\mathbf{e}_j^\top\mathbf{W}^\top))+\nonumber\\
    &\operatorname{tr}(\mathbf{I}_{nT}-\mathbf{H})(\mathbf{I}_T\otimes \mathbf{W}\mathbf{e}_j\mathbf{e}_j^\top)\mathbf{YY}^\top(\mathbf{I}_T\otimes \mathbf{e}_i\mathbf{e}_i^\top\mathbf{W}^\top)\Big]-\nonumber \\
    &T((\mathbf{I}_n-\mathbf{WP})^{-1}\mathbf{W})_{ij}((\mathbf{I}_n-\mathbf{WP})^{-1}\mathbf{W})_{ji}\label{hessrho}
\end{align}
 The deduction of \eqref{gradrho} is shown in \ref{apenA}. So, taking into account the estimation methodology of \citet{anselin1998introduction} the following estimation method is proposed for the model \eqref{PSAR}:
\begin{enumerate}
\item Assume that $\pmb{\rho}^{(0)}=\pmb{\rho}_0$.
\item With the value of $\pmb{\rho}^{(0)}$, the next fisher scoring algorithm is proposed for $\pmb{\rho}$:
\begin{equation}\label{rhoes}
    \pmb{\rho}^{(m+1)} = \pmb{\rho}^{(m)}-\mathcal{J}_{\pmb{\rho}}^{-1}(\pmb{\rho}^{(m)})\frac{\partial{\ell}}{\partial\boldsymbol{\rho}}(\pmb{\rho}^{(m)})
\end{equation}
with $\frac{\partial{\ell}}{\partial\boldsymbol{\rho}}$ defined in Equation \eqref{gradrho}, and 
\begin{align}
    \mathcal{J}_{\pmb{\rho}}&=-\left\{\mathbb{E}\left(\frac{\partial^2\ell}{\partial\rho_j \partial\rho_i}\right)\right\}_{n\times n}\label{hessianrho}
\end{align}
The exact form of $\mathcal{J}_{\pmb{\rho}}$ is unknown because the expectation of $[\mathbf{Y}^\top\mathbf{A}^\top(\mathbf{I}_{nT}-\mathbf{H})\mathbf{AY}]^{-2}$ do not have close form, but and approximation is shown in \ref{apenB}. This step is repeated until $\vert \pmb{\rho}^{(m+1)}-\pmb{\rho}^{(m)}\vert<\varepsilon$.
\item Take $\hat{\pmb{\rho}}=\pmb{\rho}^{(m+1)}$ from the previous step.
\item Calculate $\hat{\pmb{\beta}}$ and $\hat{\sigma}^2$ with $\hat{\pmb{\rho}}$ and using equations \eqref{betarho} and \eqref{sigmabetarho}, respectively.
\end{enumerate}
With the estimation method above, the following results are obtained:
\begin{theorem}\label{te1}
If the assumptions of the model defined in Equations \eqref{PSAR} are satisfied, and $n$ is fixed, then:
\begin{align}
\hat{\pmb{\rho}} & \xrightarrow[T\to \infty]{d} N_n\left(\pmb{\rho}, \mathcal{J}_{\pmb{\rho}}\right) \\
\hat {\pmb{\beta}}  & \xrightarrow[T\to \infty]{d} N_{p}\left(\pmb{\beta}, \mathbf{\sigma}^{2}(\mathbf{X}^\top\mathbf{X})^{-1}\mathbf{X}^\top\mathbf{AA}^\top\mathbf{X}(\mathbf{X}^\top\mathbf{X})^{-1} \right)
 \end{align}
\end{theorem}
\begin{proof}
Let $\rho \in (-1,1)$ and assume that $\mathbf{W}$ is row-normalized, that is, $\mathbf{W}\mathbf{1}_n = \mathbf{1}_n $, then $\mathbf{A}\mathbf{1}_{nT}=\mathbf{1}_{nT}$. For the above, $\mathbf{A}^{-1}$ exits and, $f_{\mathbf{Y}}(\mathbf{y})$ satisfies all conditions of Theorem 2.6 of \citet[pg 440]{casella2002point}; therefore, the estimators proposed in \eqref{betarho} and the obtained in the $m$-th step of Equation \eqref{rhoes} are maximum likelihood estimators. Like $n$ is fixed, but $T\to \infty$ then both estimators are normally asymptotically and unbiased, with asymptotic variance defined by:
\begin{align*}
    \lim_{T\to\infty}Var(\hat{\pmb{\rho}}) &= \mathcal{J}_{\rho}\\
    \lim_{T\to\infty}Var(\hat{\pmb{\beta}}) &= \mathbf{\sigma}^{2}(\mathbf{X}^\top\mathbf{X})^{-1}\mathbf{X}^\top\mathbf{AA}^\top\mathbf{X}(\mathbf{X}^\top\mathbf{X})^{-1}
\end{align*}
\end{proof}
Theorem 1 allows inferences to be made about the $\pmb{\beta}$ and $\pmb{\rho}$ parameters. The following Corollary is presented to decide whether it is necessary to perform a SAR model with specific coefficients for the model proposed in \eqref{PSAR}.
\begin{corollary}\label{cor1}
If the hypothesis $H_0: \rho_i=\rho, \; \forall i=1, \ldots,n$ is true, then the statistic $T^2$ defined by: 
\begin{equation}\label{tdehotelling}
T^2 = nT(\hat{\pmb{\rho}}_0-\hat{\pmb{\rho}})^\top \left[\widehat{Var}(\rho_0)\mathbf{J} +\hat{\mathcal{J}}_{\pmb{\rho}}^{-1}\right]^{-1}(\hat{\pmb{\rho}}_0-\hat{\pmb{\rho}})
\end{equation}
follows a asympotic Hotelling's $T^2$ distribution when $\hat{\pmb{\rho}}_0=(\hat{\rho}, \ldots, \hat{\rho})^\top$ is the estimator of $\rho$ obtained by a usual SAR model, and $\hat{Var}(\hat{\rho})$ is its respectively variance estimated by a usual SAR model \citep{arbia2016spatial}, $\mathbf{J}$ is a matrix of 1's of size $n\times n$, and $\hat{\pmb{\rho}}$ and $\hat{\mathcal{J}}_{\pmb{\rho}}$ are obtained by the methodology proposed in this work. Moreover, $\frac{2nT-n-1}{n(2nT-1)}T^2 \approx F_{(n,v)}$ with
\begin{equation}\nonumber
\frac{1}{v}= \frac{1}{nT-1}\sum_{i=1}^2\left(\frac{(\hat{\pmb{\rho}}_0-\hat{\pmb{\rho}})^\top \mathbf{S}^{-1}\left(\frac{1}{nT}\mathbf{S}_i\right)\mathbf{S}^{-1}(\hat{\pmb{\rho}}_0-\hat{\pmb{\rho}})}{T^2}\right)^2
\end{equation}
where $\mathbf{S}^{-1}=\frac{1}{nT}(\widehat{Var}(\rho_0)\mathbf{J} +\hat{\mathcal{J}}_{\pmb{\rho}}^{-1})$, $\mathbf{S}_1=\widehat{Var}(\rho_0)\mathbf{J}$ and $\mathbf{S}_2= \hat{\mathcal{J}}_{\pmb{\rho}}^{-1}$.
\end{corollary}
\begin{proof}
If $\rho_i=\rho$, $\forall i$, then Equation \eqref{PSAR} becomes:
\begin{align}
    {Y}_{it}&=\rho\sum_{j=1}^n w_{ij}y_{jt} + \sum_{k=1}^K \beta_{k} x_{itk}  + \epsilon_{it}\nonumber
\end{align}
and let $\mathbf{B}=\mathbf{I}_{nT}-(\mathbf{I}_T\otimes \rho\mathbf{W})=\mathbf{I}_{T}\otimes (\mathbf{I}_n-\rho\mathbf{W})=\mathbf{I}_{nT}-\rho\Tilde{\mathbf{W}}$, with $\Tilde{\mathbf{W}}=\mathbf{I}_T\otimes\mathbf{W}$ and $\hat{\rho}$ obtained by Equation:
\begin{align*}
    \hat{\rho}&=\arg \max\left\{ -\frac{nT}{2}\ln(2\pi\sigma^2)+\ln|\mathbf{B}|-\frac{1}{2\sigma^2}(\mathbf{By}-\mathbf{X}\pmb{\beta})^\top
(\mathbf{By}-\mathbf{X}\pmb{\beta})\right\}
\end{align*}
As $\Tilde{W}$ is row-normalized, then $\hat{\rho} \xrightarrow[T\to \infty]{d} N\left(\rho, Var(\hat{\rho})\right)$ \citep{arbia2016spatial}. Let $\hat{\pmb{\rho}}_0=(\hat{\rho}, \ldots, \hat{\rho})^\top$ then, by Equation (8) of \citet{kumar2014eigenvalue} is true that:
\begin{equation*}
    nT(\widehat{Var}(\rho_0)\mathbf{J}+ \hat{\mathcal{J}}_{\pmb{\rho}}^{-1})^{-1}\xrightarrow[T\to \infty]{d}\mathcal{W}(nT-n-1, Var(\hat{\rho}_0)\mathbf{J}+ {\mathcal{J}}_{\pmb{\rho}}^{-1})
\end{equation*}
where $\mathcal{W}$ is a Wishart distribution. Then, the statistic $T^2$ converges to a Hotelling's $T^2$ distribution \citep{anderson1958introduction, BARINGHAUS2017177}. 
Finally using Theorem 2 of \citet{pan2011central}, the form of $v$ is obtained.
\end{proof}

The estimation algorithm proposed in this section is computationally complex due to the sizes $n$, $T$, and $K$. Furthermore, being a Fisher scoring algorithm, at each step it is necessary to calculate products of matrices of size $nT\times nT$ and inverses of size $nK\times nK$. Therefore, a good selection of the initial value $\pmb{\rho}_0$ is crucial.

Furthermore, two important assumptions of the model: normality and homoscedasticity in the innovations make the application of the model in various scenarios very restrictive. To solve these two drawbacks, an estimator is constructed that is the extension of the estimator proposed by \cite{KELEJIAN201053} for a traditional SAR with heteroskedastic innovations.

\subsection{Heteroscesdatic SAR models with specific spatial coefficients}\label{hetSPSAR}
The estimator proposed by \cite{KELEJIAN201053} is useful in SAR with a single coefficient. Adapting the assumptions and constructing appropriate matrices, the following theorem gives us the robust estimator of the model defined in Equation \eqref{PSAR}. 
\begin{theorem}\label{the2}
If the next assumptions are true:
\begin{enumerate}
\item The model proposed in Equation \eqref{PSAR} is true.
\item $\mathbf{W}\mathbf{1}_n =\mathbf{1}_n$. 
\item The matrix $\mathbf{D}$ defined by:
\begin{equation}\nonumber
  \mathbf{D} =\begin{bmatrix}
    y_{11}\mathbf{W}_{\cdot 1} &  y_{21}\mathbf{W}_{\cdot 2}&\cdots &  y_{n1}\mathbf{W}_{\cdot n}\\
    y_{12}\mathbf{W}_{\cdot 1} &  y_{22}\mathbf{W}_{\cdot 2}&\cdots &  y_{n2}\mathbf{W}_{\cdot n}\\
    \vdots& \vdots&\ddots &  \vdots \\
    y_{1T}\mathbf{W}_{\cdot 1} &  y_{2T}\mathbf{W}_{\cdot 2}&\cdots &  y_{nT}\mathbf{W}_{\cdot n}
\end{bmatrix}
\end{equation}
when $\mathbf{W}_{\cdot i}$ is the $i$-th column of the matrix $\mathbf{W}$,
satisfies that are uniformly bounded in absolute value.
\item The innovations $\epsilon_{it}$ in Equation \eqref{PSAR} satisfy $\mathbb{E}(\epsilon_{it})=0$ and $Var(\epsilon_{it})=\sigma_{it}^2<\infty$, and $\sup_{n,T}(\mathbb{E}(\epsilon_{it}^{4+\eta}))<\infty$, $\exists \eta>0$ and $\epsilon_{it}$ is independent of $\epsilon_{i't'}$ $\forall i\neq i' \,or\, t\neq t' $.
\item The matrix $\mathbf{H}$ conformed by the independent columns of the matrix $[\mathbf{X}, (\mathbf{1}_{T\times T}\otimes \mathbf{W})\mathbf{X}, (\mathbf{1}_{T\times T}\otimes \mathbf{W}^2)\mathbf{X}, \ldots, (\mathbf{1}_{T\times T}\otimes \mathbf{W}^q)\mathbf{X}]$ satisfies $rank(\mathbf{H})>rank(\mathbf{X})$ for some $q\geq 2$. 
\item The matrix $\mathbf{Q_{HH}}=\lim_{T\to\infty}\frac{1}{T}\mathbf{H}^\top\mathbf{H}$ is finite and non-singular.
\item The matrix $\mathbf{Q_{HZ}}=\lim_{T\to\infty}\frac{1}{T}\mathbf{H}^\top\mathbf{Z}$ is finite and non-singular with $\mathbf{Z}=[\mathbf{X}, \mathbf{P_HD}]$.
\end{enumerate}
then the estimator of $\pmb{\delta}=(\pmb{\beta}^\top, \pmb{\rho}^\top)^\top$ proposed as:
\begin{equation}\label{kj20}
    \hat{\pmb{\delta}}=\left[(\mathbf{P_H}\mathbf{Z})^\top\mathbf{Z}\right]^{-1}(\mathbf{P_H}\mathbf{Z})^\top\mathbf{Y}
\end{equation}
where $\mathbf{P_H}=\mathbf{H}(\mathbf{H}^\top\mathbf{H})^{-1}\mathbf{H}^\top$, $\mathbf{Z}=[\mathbf{X}, \mathbf{P_HWY}]$ satisfies that:
\begin{enumerate}
    \item $T^{\frac{1}{2}}(\hat{\pmb{\delta}}-\pmb{\delta})=T^{\frac{1}{2}}\mathbf{B}\pmb{\epsilon} + o_{n+K}(1)$ with $\mathbf{B}=\mathbf{Q_{HH}}^{-1}\mathbf{Q_{HZ}}^\top[\mathbf{\mathbf{Q_{HZ}}^\top Q_{HH}}^{-1}\mathbf{Q_{HZ}}]^{-1}\mathbf{H}^\top$
    \item $T^{\frac{1}{2}}\mathbf{B}^\top\pmb{\epsilon}=\mathcal{O}_{n+K}(1)$
    \item $\hat{\pmb{\delta}}\xrightarrow[T\to \infty]{d} N_{K+n}\left(\pmb{\delta}, \frac{1}{T}\mathbf{B}^\top\mathbf{B}\right)$
\end{enumerate}
\end{theorem}
\begin{proof}
See \ref{apenC}
\end{proof}
This theorem provides us with several very important results. From a theoretical point of view, assumption 3 of the theorem is less restrictive than the normal and homoscedastic models. It only requires innovations with zero mean and finite fourth moment. In addition, the estimator gives us the asymptotic form of the variance of the estimators, which allows classical inference. In addition, this robust estimator can be used as an initial value for the estimation procedure of Equation \eqref{rhoes}. In simulation scenarios, this initial value significantly increases the speed of convergence when $n$ is large. 
Comparing the performance of the two estimators, the following corollary is obtained:
\begin{corollary}\label{cor2}
Let $\pmb{\delta}=(\pmb{\beta}^\top, \pmb{\rho}^\top)^\top$, $\hat{\pmb{\delta}}_{Hom}=(\hat{\pmb{\beta}}^\top, \hat{\pmb{\rho}}^\top)^\top$ obtained by equations \eqref{betarho} and \eqref{rhoes} respectively, and  $\hat{\pmb{\delta}}_{Het}$ the estimator obtained by Equation \eqref{kj20}.  If the model proposed in Equation \eqref{PSAR} is true, and $\epsilon_{it}\sim N(0, \sigma^2)$ then:
\begin{equation}\nonumber
\lim_{T\to \infty}\left(\mathrm{E}[(\hat{\pmb{\delta}}_{Hom}-\pmb{\delta})^2]-\mathrm{E}[(\hat{\pmb{\delta}}_{Het}-\pmb{\delta})^2]\right)< 0 \qquad a.s.\; \pmb{\delta}\in \mathbb{R}^{K}\times (-1,1)^n
\end{equation}
\end{corollary}
\begin{proof}
By Theorem \ref{te1} and Theorem 2.6 of \cite[pg 440]{casella2002point}, it is clear that maximum likelihood estimator $\hat{\pmb{\delta}}_{Hom}$  reaches the  Cramer-Rao bound, so it is true that:
\begin{equation}\nonumber
  \lim_{T\to \infty}\left(\mathrm{E}[(\hat{\pmb{\delta}}_{Hom}-\pmb{\delta})^2]-\mathrm{E}[(\hat{\pmb{\delta}}_{Het}-\pmb{\delta})^2]\right)\leq 0 \qquad a.s.\; \pmb{\delta}\in \mathbb{R}^{K}\times (-1,1)^n
\end{equation}
Since Equation \eqref{rhoes} is not a polynomial function, then by Taylor's Theorem over measurable functions \citep[pp 87–135]{jiang2010large}, it is found that:
\begin{equation}\nonumber
    \hat{\pmb{\delta}}_{Hom}\neq \hat{\pmb{\delta}}_{Het} \qquad a.s.\; \pmb{\delta}\in \mathbb{R}^{K}\times (-1,1)^n
\end{equation}
therefore, the robust estimator obtained in Equation \eqref{kj20} will almost sure not reach the Cramer-Rao bound.
\end{proof}
This corollary guarantees that if the assumptions of homoscedasticity and normality are true, the estimator obtained by the maximum likelihood method is better since it will have a lower mean square error and lower bias.
\section{Simulation}
Two simulation studies were conducted to evaluate the proposed methodology: one assuming normal distribution with homoscedastic innovations and the other with non-normal and heteroscedastic innovations. Each simulation scenario was repeated 500 times. The R codes \citep{Rmanual} are available in the supplementary file \ref{sf1}.  To assess performance against the original simulation parameters, the following models were adjusted:
i) Ho-SAR: Maximum likelihood SAR with homoscedasticity \citep{anselin1988spatial},
ii) Ho-Proposed: PSAR with homoscedasticity (proposed methodology in Section \ref{hoSPSAR}), iii) Ro-Proposed: PSAR with heteroscedasticity (proposed methodology in Section \ref{hetSPSAR}), iv) Ro-SAR: Maximum likelihood SAR robust to heteroscedasticity \citep{arraiz2010spatial}, and v) Ro-SARAR: SARAR model with robust estimation for heteroscedastic perturbations \citep{arraiz2010spatial}.

\subsection{Homoscedastic innovations}
Each $\pmb{y}=(\pmb{y}_1, \ldots, \pmb{y}_T)$ generated from a normal distribution on a regular grid was simulated, with $\pmb{\epsilon} \sim N(\pmb{0}, \mathbf{I}_{nT\times nT})$, $\pmb{y}=\left(\mathbf{I}_{nT}-(\mathbf{I}_T\otimes \mathbf{WP} )\right)^{-1}[\mathbf{X}\pmb{\beta}+\pmb{\epsilon}]$, where $x_{1i}\sim N(0,1)$, $x_{2i}\sim N(2,1)$, $\mathbf{X}=(\mathbf{1}_{nT}, \mathbf{x}_1, \mathbf{x}_2)$ and the $w_{ij}$ follow a tower-like first-order contiguity, $i=1, \ldots, n$, $n=25, 49, 64, 81, 144$, $T=5,10, 50, 100,200$, each value of $\rho_i\in -0.9, -0.8,\ldots, 0.8, 0.9$, $\beta_0=1$, $\beta_1=-1$, $\beta_2=0.5$. 
\begin{figure}[!ht]
\centering
\includegraphics[width=14cm]{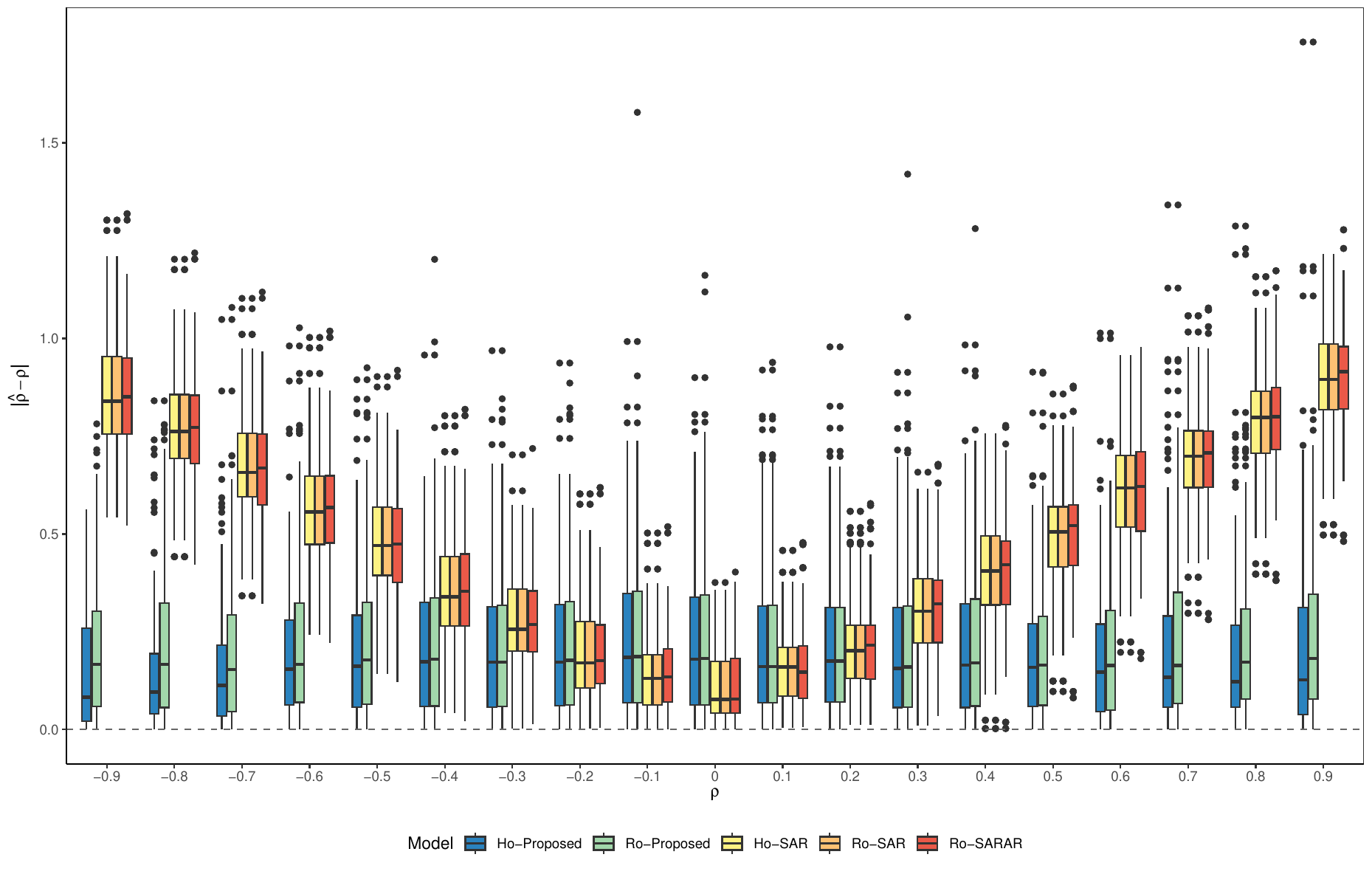}
\caption{Boxplots of $\hat{\rho}_i$ when $T=10$, $\beta_0=0.5$, $\beta_1=-0.5$, $\beta_2=1$ and $n=49$.}
\label{rho4910}
\end{figure}

Figure \ref{rho4910} presents the value $\vert \hat{\rho}_i - \rho_i \vert $ for the estimators obtained using the methodology described in subsection \ref{hoSPSAR} (referred to as Ho-Proposed) and the one in subsection \ref{hetSPSAR} (referred to as Ro-Proposed). In this figure, the spatial domain consists of $n = 49$ regions, arranged in a $7 \times 7$ grid, with each region observed $T = 10$ times.
The results indicate that the estimator assuming homoscedasticity exhibits lower variance compared to the heteroscedastic counterpart. This aligns with the conclusions of Theorem \ref{te1} and Corollary \ref{cor2}, given that the assumptions of normality and homoscedasticity hold. However, both estimators show high variability, mainly due to the limited sample size ($T = 10$), which is insufficient for obtaining consistent and efficient estimates of each $\rho_i$.
Additionally, the three alternative models that assume a constant spatial correlation $\hat{\rho}_i = \hat{\rho}$ exhibit a significant bias and fail to capture the spatial dependence structure accurately.

\begin{figure}[!ht]
\centering
\includegraphics[width=14cm]{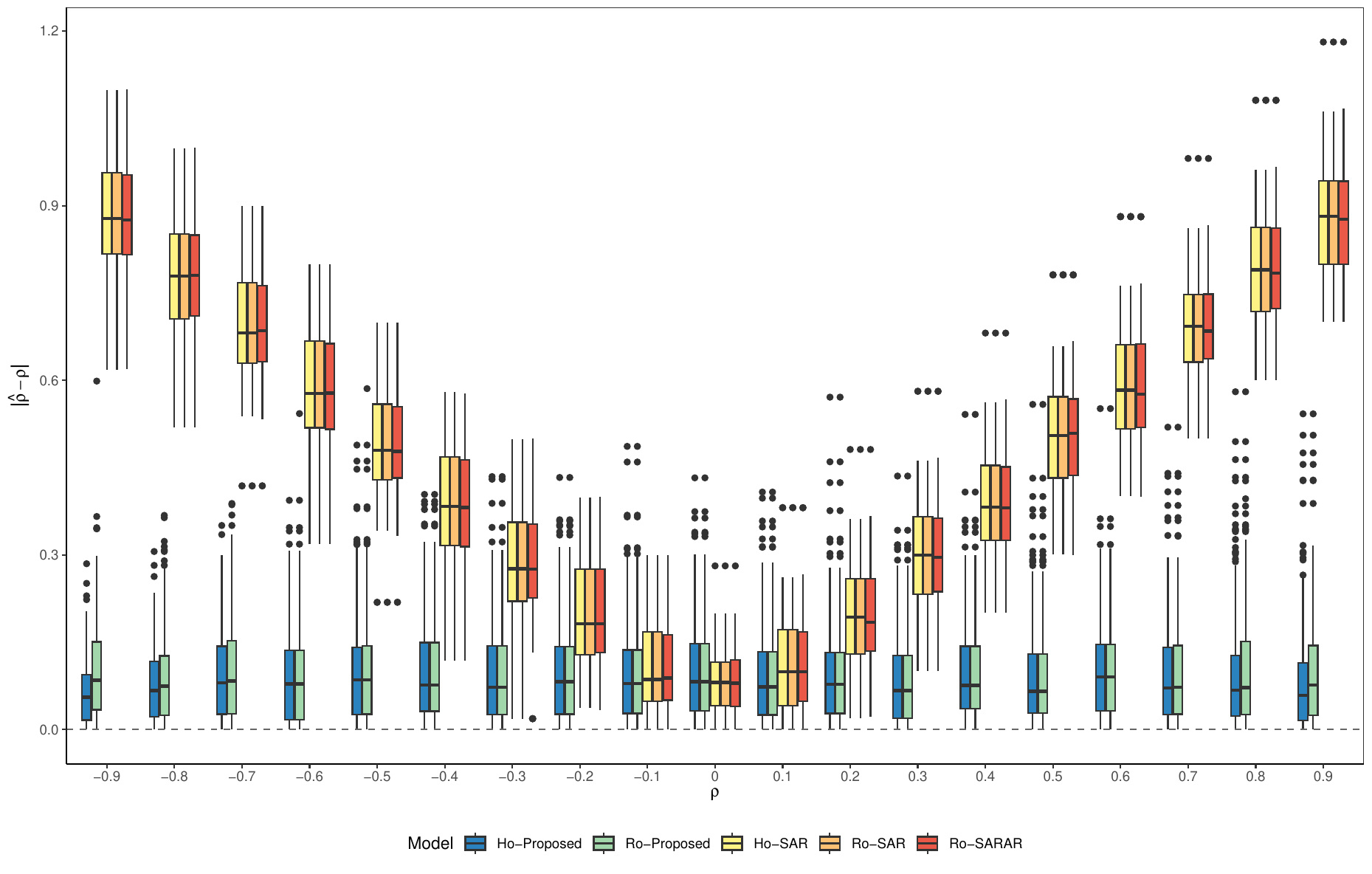}
\caption{Boxplots of $\hat{\rho}_i$ when $T=50$, $\beta_0=0.5$, $\beta_1=-0.5$, $\beta_2=1$ and $n=49$.}
\label{rho4950}
\end{figure}

Figure \eqref{rho4950} displays the estimator of $ \rho_i $ obtained using the methodology from section \ref{hoSPSAR}, referred to as Ho-Proposal, and the one from section \ref{hetSPSAR}, called Ro-Proposal. The spatial domain remains $ n=49 $, as in Figure \ref{rho4910}, but each region is now observed $ T=50 $ times.
It is evident that the estimator under the homoscedasticity assumption has a lower variance compared to the heteroscedastic case in line with Corollary \ref{cor2}. Moreover, its variance is significantly reduced compared to when $ T=10 $. Simulations with different temporal sample sizes further confirm the consistency of estimators, aligning with the results of Theorem \ref{te1}. 

\begin{figure}[!ht]
\centering
\includegraphics[width=14cm]{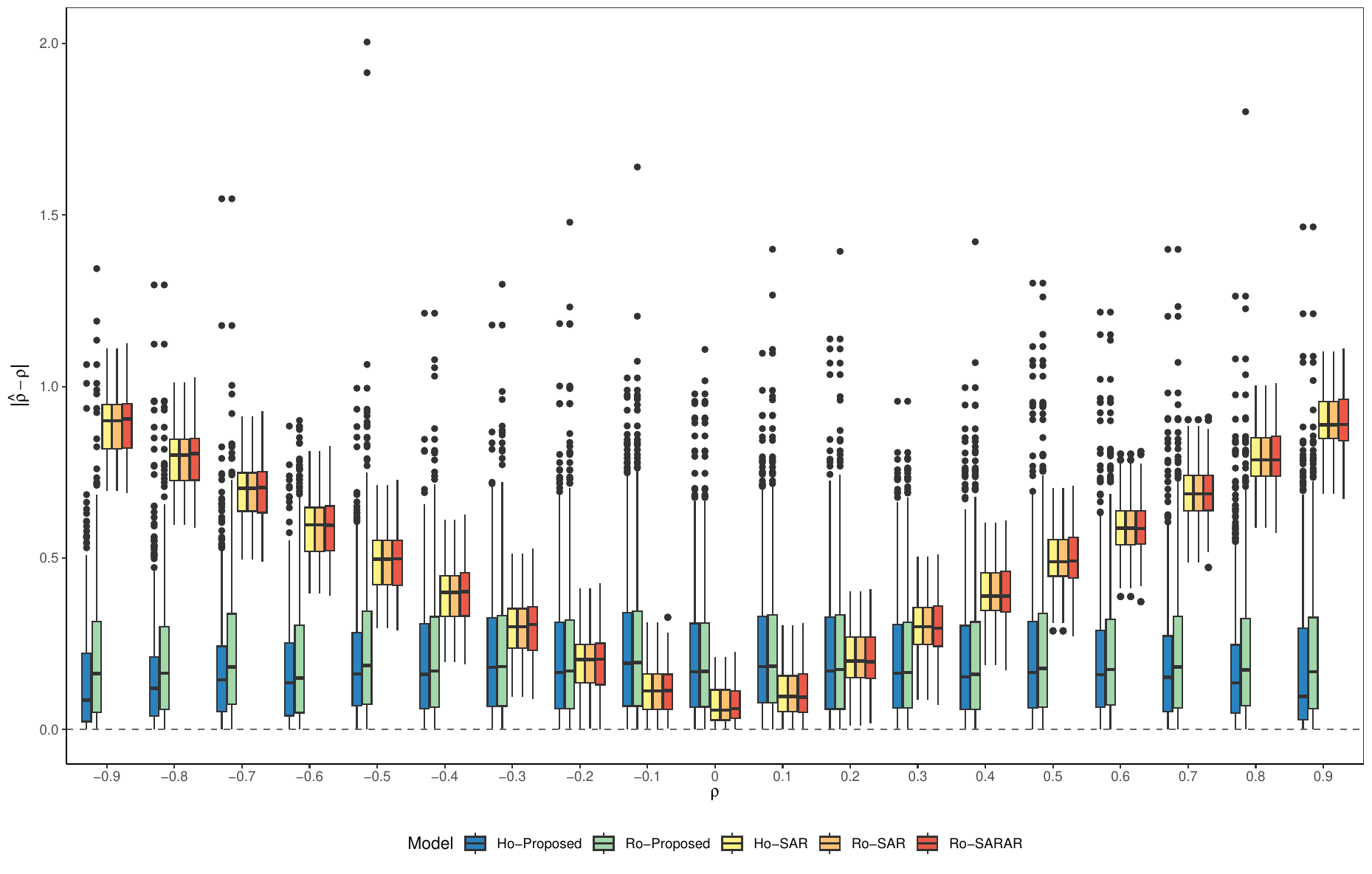}
\caption{Boxplots of $\hat{\rho}_i$ when $T=10$, $\beta_0=0.5$, $\beta_1=-0.5$, $\beta_2=1$ and $n=144$.}
\label{rho14410}
\end{figure}

Furthermore, the three alternative models that assume a uniform spatial correlation ($ \hat{\rho}_i = \hat{\rho} $) exhibit substantial bias and fail to adequately capture spatial dependence. Notably, these models consistently produce estimates close to zero, implying conclusions based on the absence of spatial correlation. In contrast, when allowing for multiple $ \rho_i $, the estimator remains centered at zero. It is important to highlight that these models are not intended to estimate specific coefficients but are included in the simulation to facilitate comparison of the covariate effects, represented by the vector $\pmb{\beta}$.
\begin{figure}[!ht]
\centering
\includegraphics[width=14cm]{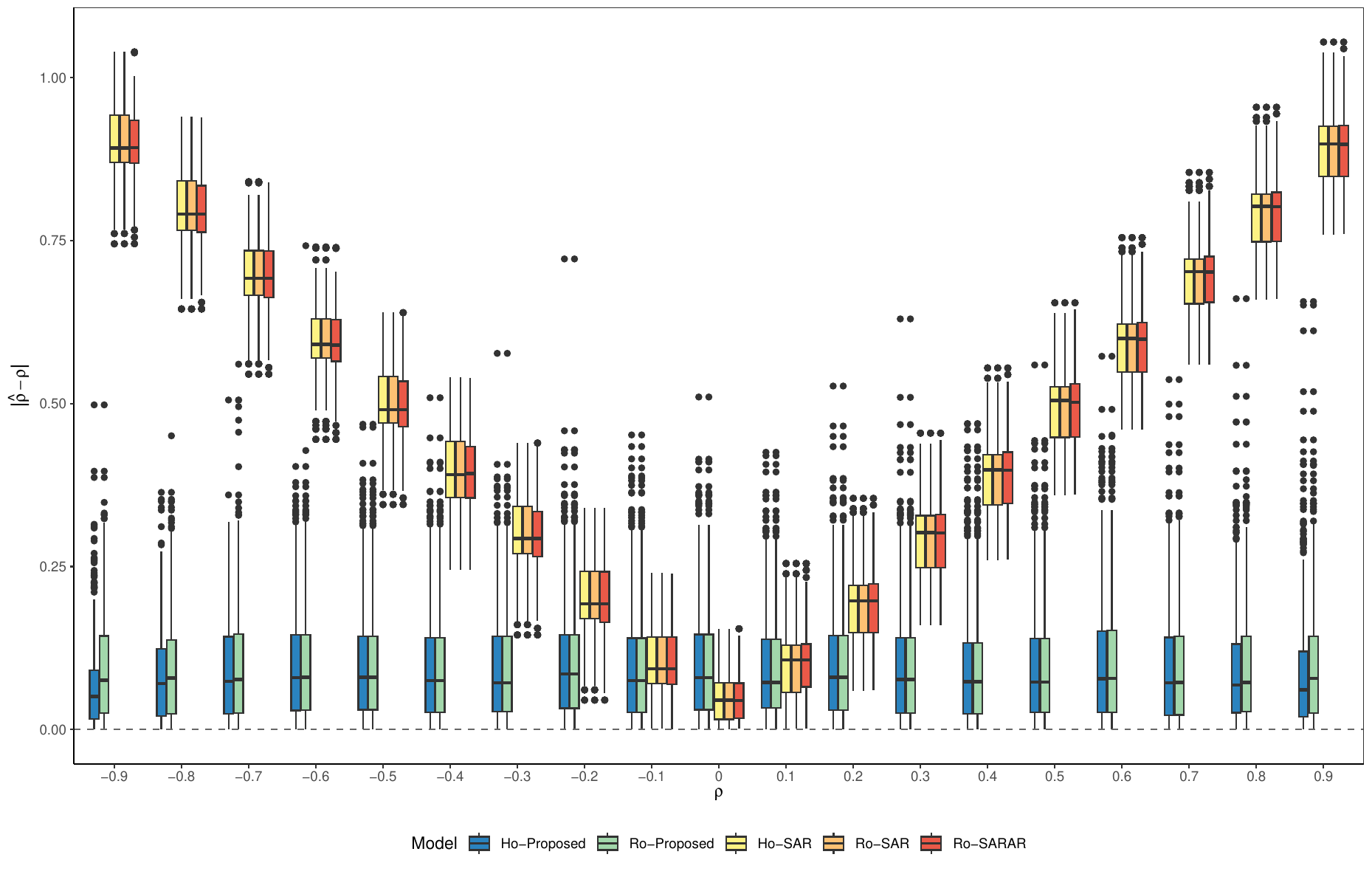}
\caption{Boxplots of $\hat{\rho}_i$ when $T=50$, $\beta_0=0.5$, $\beta_1=-0.5$, $\beta_2=1$ and $n=144$.}
\label{rho14450}
\end{figure}

Furthermore, it is noteworthy that the bias of the estimator becomes very small, except for values of $\rho_i$ close to 1, similar to what occurs in standard SAR models \citep{arbia2016spatial, lesage2009introduction}.  
Although the variance of the estimator proposed in section \ref{hetSPSAR} also decreases, some outliers remain, highlighting the need for a larger $T$ when dealing with heteroscedastic data, as discussed in the simulation studies of \citet{shaker2015spatial}.  
Figure \ref{rho14450} presents the same analysis as Figure \ref{rho4950} but for a spatial size of $n=144$. The results are largely similar, suggesting that the consistency of $\rho_i$ estimation depends more on $T$ than on $n$.  

\begin{figure}[!ht]
\centering
\includegraphics[width=14cm]{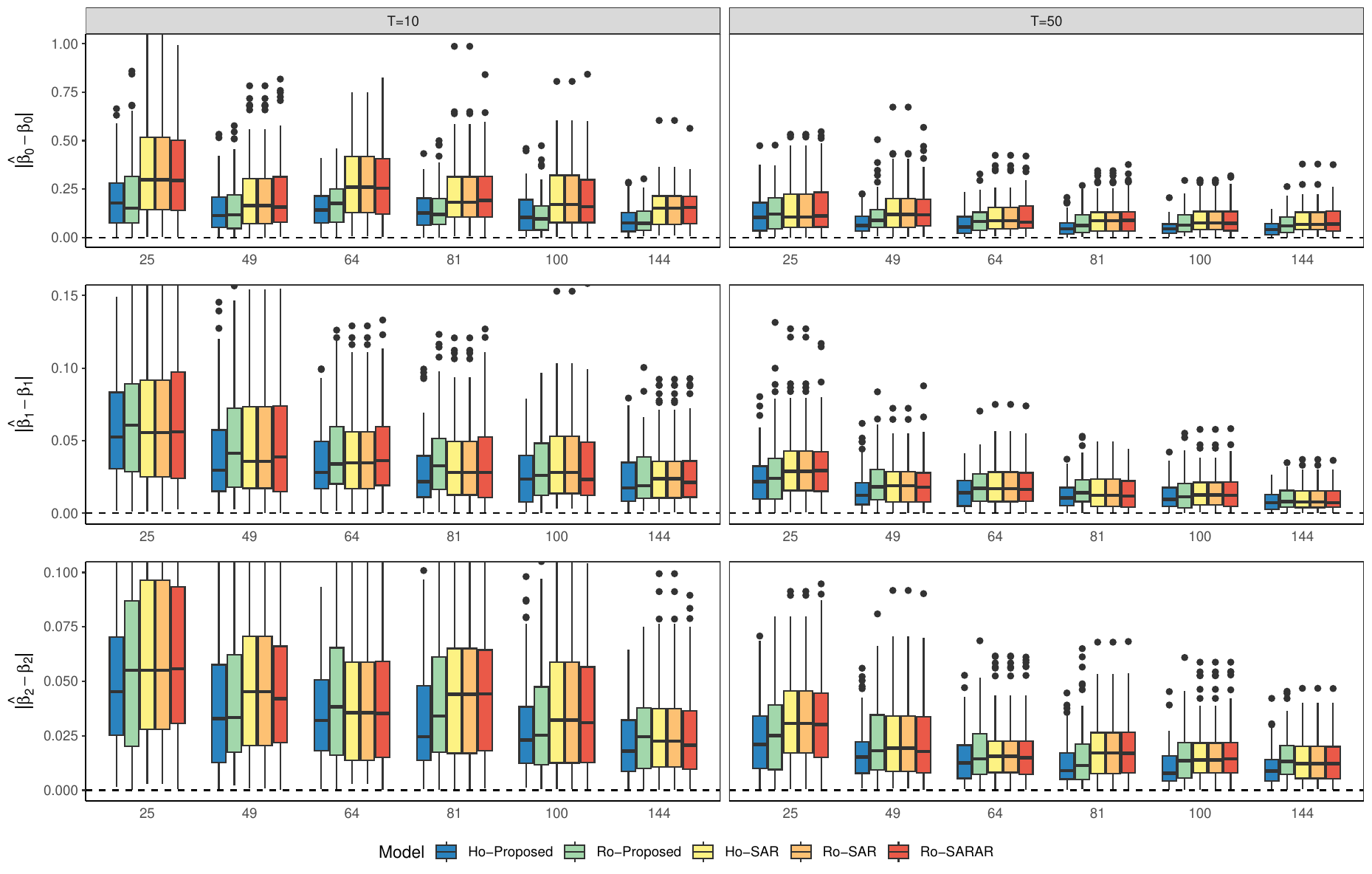}
\caption{Boxplots of $\hat{\beta}_k$ when $T=10,50$, $\beta_0=0.5$, $\beta_1=-0.5$, $\beta_2=1$ and $n=25, 49, 64, 81, 100, 144$.}
\label{betaHo1050}
\end{figure}

\begin{figure}[H]
\centering
\includegraphics[width=14cm]{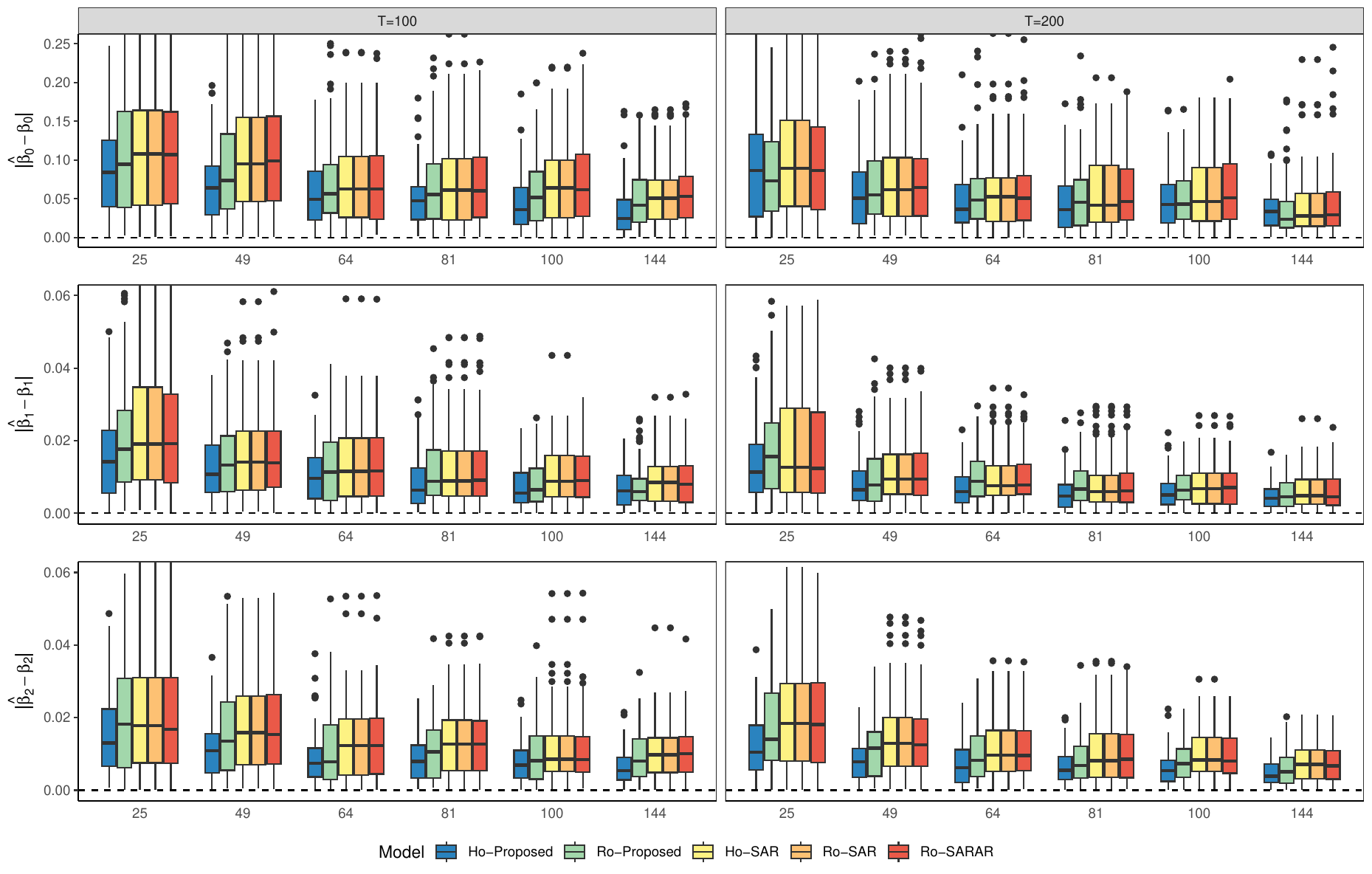}
\caption{Boxplots of $\hat{\beta}_k$ when $T=100, 200$, $\beta_0=0.5$, $\beta_1=-0.5$, $\beta_2=1$ and $n=25, 49, 64, 81, 100, 144$.}
\label{betaHo100200}
\end{figure}

Figure \ref{betaHo1050} presents the absolute error $\vert \hat{\beta}_k - \beta_k \vert$ for each $\beta_k$ across the five proposed models when the number of temporal observations is 10 and 50, respectively. Notably, the Ho-Proposed model performs best when $T = 50$, with its accuracy improving further as $n$ increases, followed by the Ro-Proposed model. The other three models yield estimators with minimal bias since the covariates were generated independently of the spatial grid.  
Moreover, in all cases, the theoretical results ensure the asymptotic reliability of the estimators for $\pmb{\beta}$.  

Figure \ref{betaHo100200} displays $\vert \hat{\beta}_k - \beta_k \vert$ for the same models when $T = 100$ and $T = 200$. The patterns observed are consistent with those in Figure \ref{betaHo1050}, though the variance in each model further decreases, becoming particularly small in the Ho-Proposed model. This highlights the strong estimation performance of the proposed methodology when its assumptions hold. Additionally, the Ro-Proposed estimator remains competitive, performing well compared to conventional models assuming normality.  
These results confirm the consistency of the proposed estimator under normality and homoscedasticity conditions.  

\begin{figure}[!ht]
\centering
\includegraphics[width=14cm]{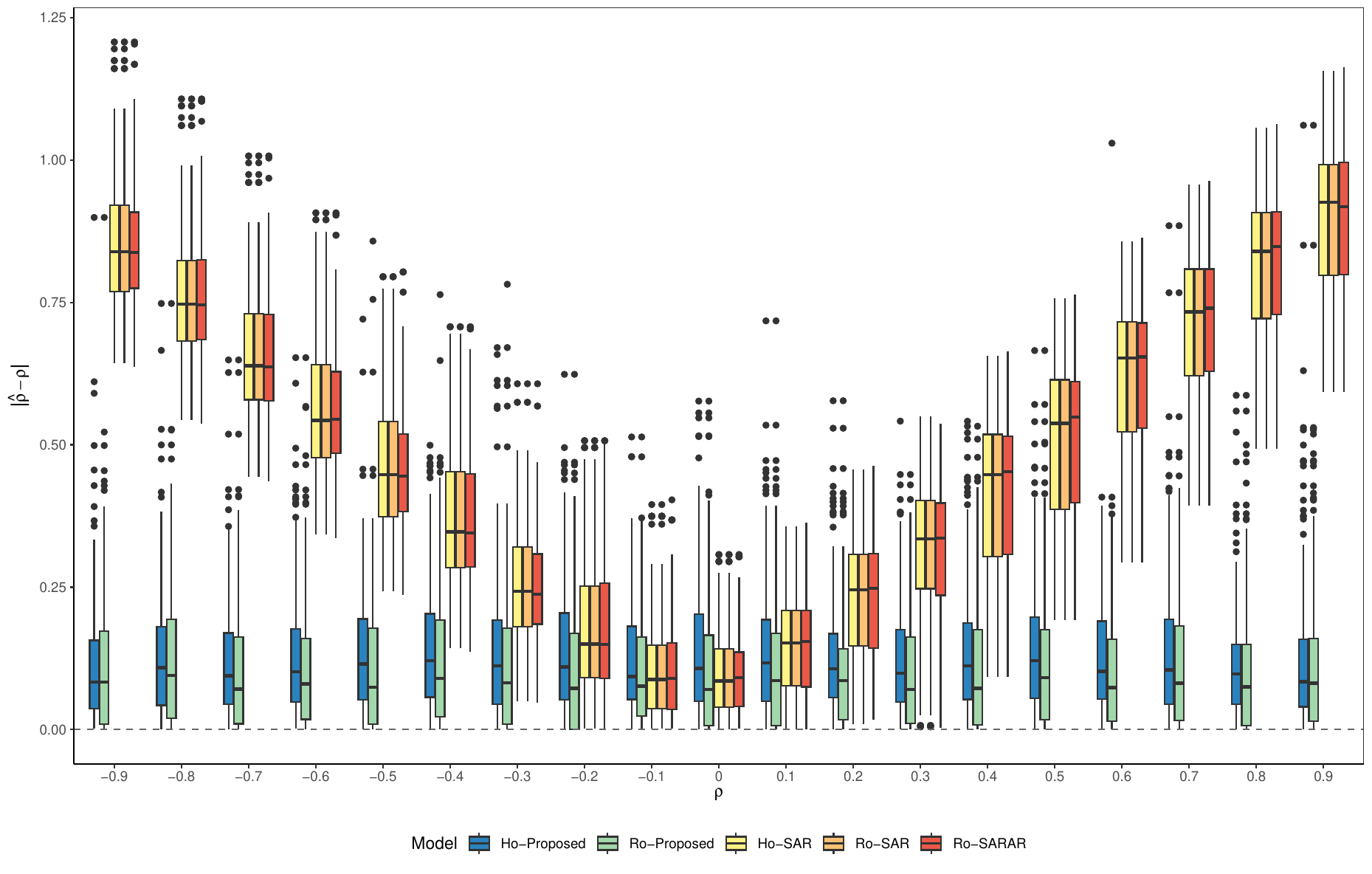}
\caption{Boxplots of $\hat{\rho}_i$ when $T=10$, $\beta_0=0.5$, $\beta_1=-0.5$, $\beta_2=1$ and $n=49$.}
\label{Hetrho4910}
\end{figure}
\subsection{Heteroskedastic innovations}
Each $\pmb{y}=(\pmb{y}_1, \ldots, \pmb{y}_T)$ generated from a non-normal distribution on a regular grid was simulated, with $\epsilon_{it}=u_{it}-\frac{1}{2}$, ${u}_{it} \sim \Gamma(v_{it},2v_{it})$, i.e. $v_{it}\sim U(0.5,1.5)$.
$$\mathrm{E}(\epsilon_{it})=0, \;Var(\epsilon_{it})=\frac{v_{it}}{4v_{it}^2}$$
Therefore, $\pmb{y}=\left(\mathbf{I}_{nT}-(\mathbf{I}_T\otimes \mathbf{WP} )\right)^{-1}[\mathbf{X}\pmb{\beta}+\pmb{\epsilon}]$, where $x_{1i}\sim N(0,1)$, $x_{2i}\sim N(2,1)$, $\mathbf{X}=(\mathbf{1}_{nT}, \mathbf{x}_1, \mathbf{x}_2)$ and the $w_{ij}$ follow a tower-like first-order contiguity, $i=1, \ldots, n$, $n=49, 81, 144, 400$, $T=5,10, 50, 100,1000$ each value of $\rho_i\sim U(-1,1)$, $\beta_0=1$, $\beta_1=-1$, $\beta_2=0.5$. Figure \eqref{Hetrho4910} presents the absolute error $\vert \hat{\rho}_i - \rho_i \vert$ when $\epsilon_{it} = u_{it} - \frac{1}{2}$, with ${u}_{it} \sim \Gamma(v_{it}, 2v_{it})$ and $v_{it} \sim U(0.5,1.5)$. This figure is analogous to Figure \ref{rho4910}.  

\begin{figure}[!ht]
\centering
\includegraphics[width=14cm]{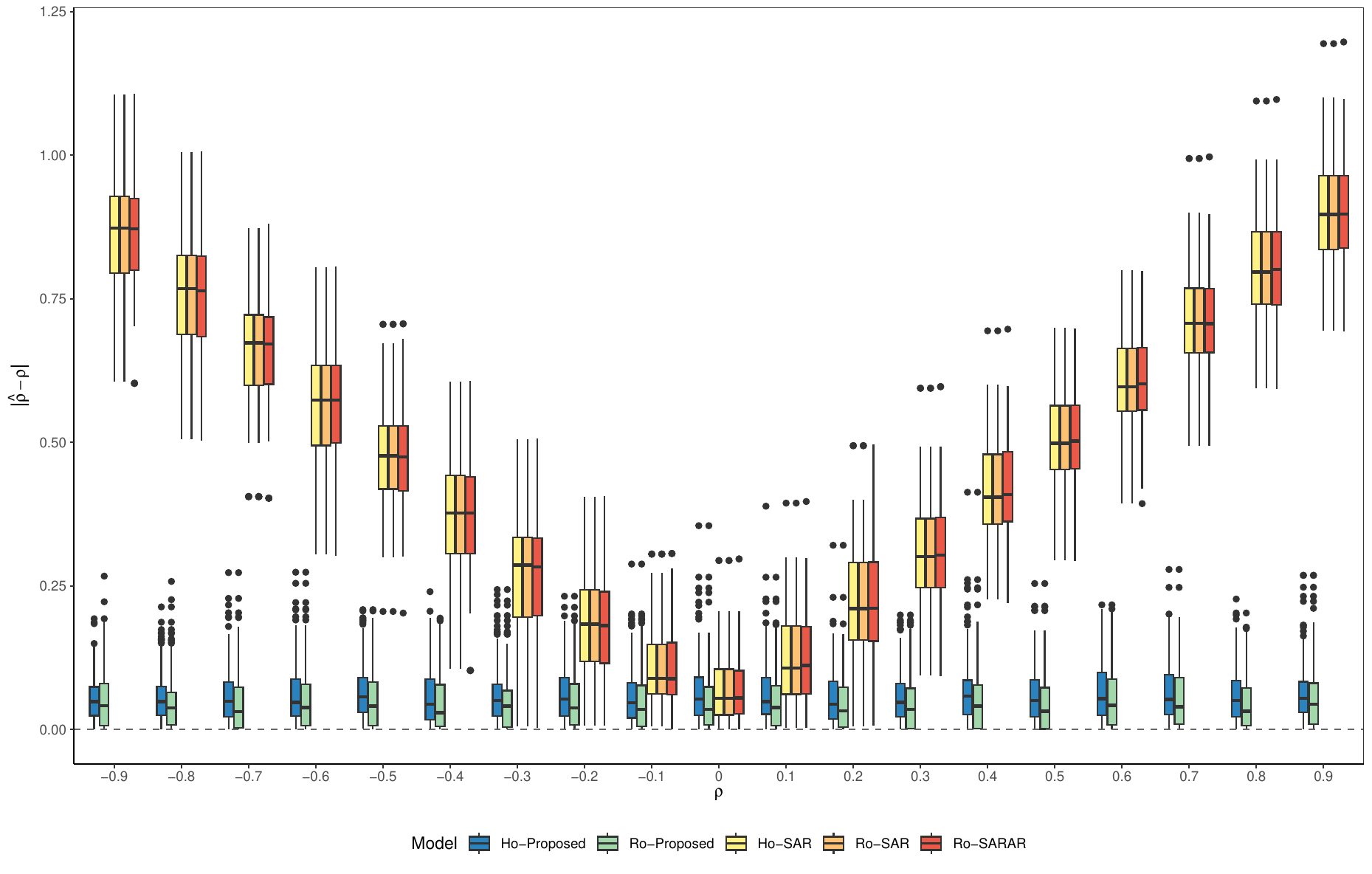}
\caption{Boxplots of $\hat{\rho}_i$ when $T=50$, $\beta_0=0.5$, $\beta_1=-0.5$, $\beta_2=1$ and $n=49$.}
\label{Hetrho4950}
\end{figure}
It is observed that the estimator assuming homoscedasticity exhibits similar variance to the heteroscedastic case but with greater bias. Notably, the proposed robust methodology does not enforce the constraint that the parameter must lie within the interval $(-1,1)$. As a result, for small sample sizes, it may produce estimates outside this range.  Additionally, the estimators show high variability, primarily because $T=10$ is too small to obtain consistent and efficient estimates for each $\rho_i$.  

Figure \eqref{Hetrho4950} presents the absolute error $\vert \hat{\rho}_i - \rho_i \vert$ when $\epsilon_{it} = u_{it} - \frac{1}{2}$, with ${u}_{it} \sim \Gamma(v_{it}, 2v_{it})$ and $v_{it} \sim U(0.5, 1.5)$.  It can be observed that the estimator assuming homoscedasticity has greater variance compared to the heteroscedastic estimator, but with a higher bias, although both models show a small bias overall. It is important to note that the proposed robust methodology may produce estimates outside the valid range when $\rho$ is very close to 1 or less than -1. Comparing the two figures, it is evident that the estimator is consistent.  

For the other models assuming $\rho_i = \rho$, the performance is poor, similar to the results observed in the homoscedastic simulations, highlighting the advantages of the two proposed estimators under homoscedasticity and heteroscedasticity in the residuals, respectively.  

\begin{figure}[!ht]
\centering
\includegraphics[width=14cm]{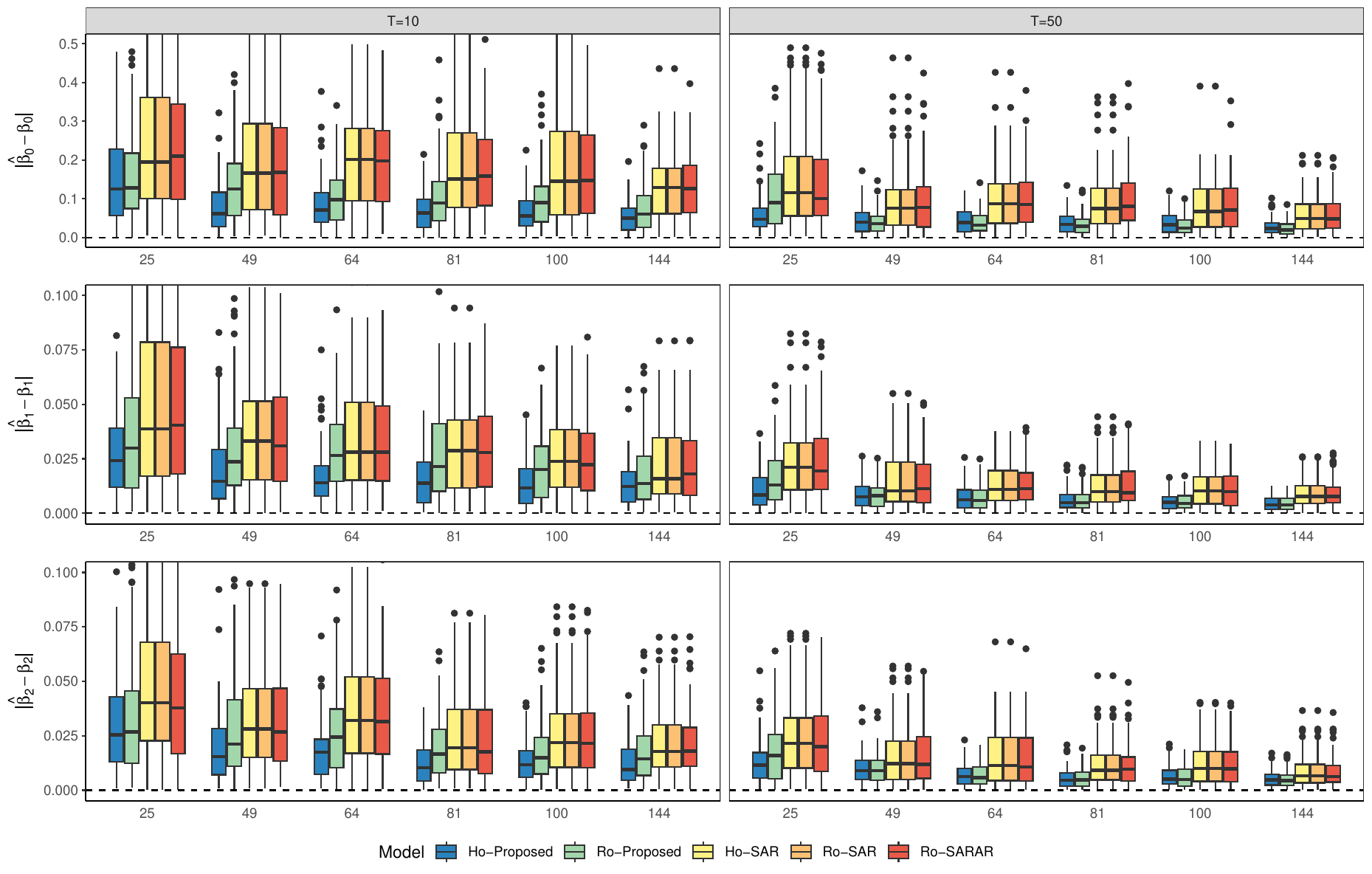}
\caption{Boxplots of $\hat{\beta}_k$ when $T=10,50$, $\beta_0=0.5$, $\beta_1=-0.5$, $\beta_2=1$ and $n=25, 49, 64, 81, 100, 144$. $\epsilon_{it} = u_{it} - \frac{1}{2}$, with ${u}_{it} \sim \Gamma(v_{it}, 2v_{it})$ and $v_{it} \sim U(0.5, 1.5)$.}
\label{betaHet1050}
\end{figure}
In addition, Figure \ref{betaHet1050} presents the results of estimating $\pmb{\beta}$ under heteroscedastic and non-normal innovations. The homoscedastic model yields good results, but the Robust model outperforms it when $n \geq 49$, and it performs significantly better than the other models. It is also noteworthy that for small $T$, the homoscedastic model still provides satisfactory results. All the simulation scenarios are presented in Supplementary File \ref{sf1}.
This simulation supports the theoretical results of Theorems \ref{te1} and \ref{the2}, and Corollary \ref{cor2}. In the following section, the methodology is applied to real data.

\section{Application}
\begin{figure}[!ht]
\centering
\includegraphics[width=14cm]{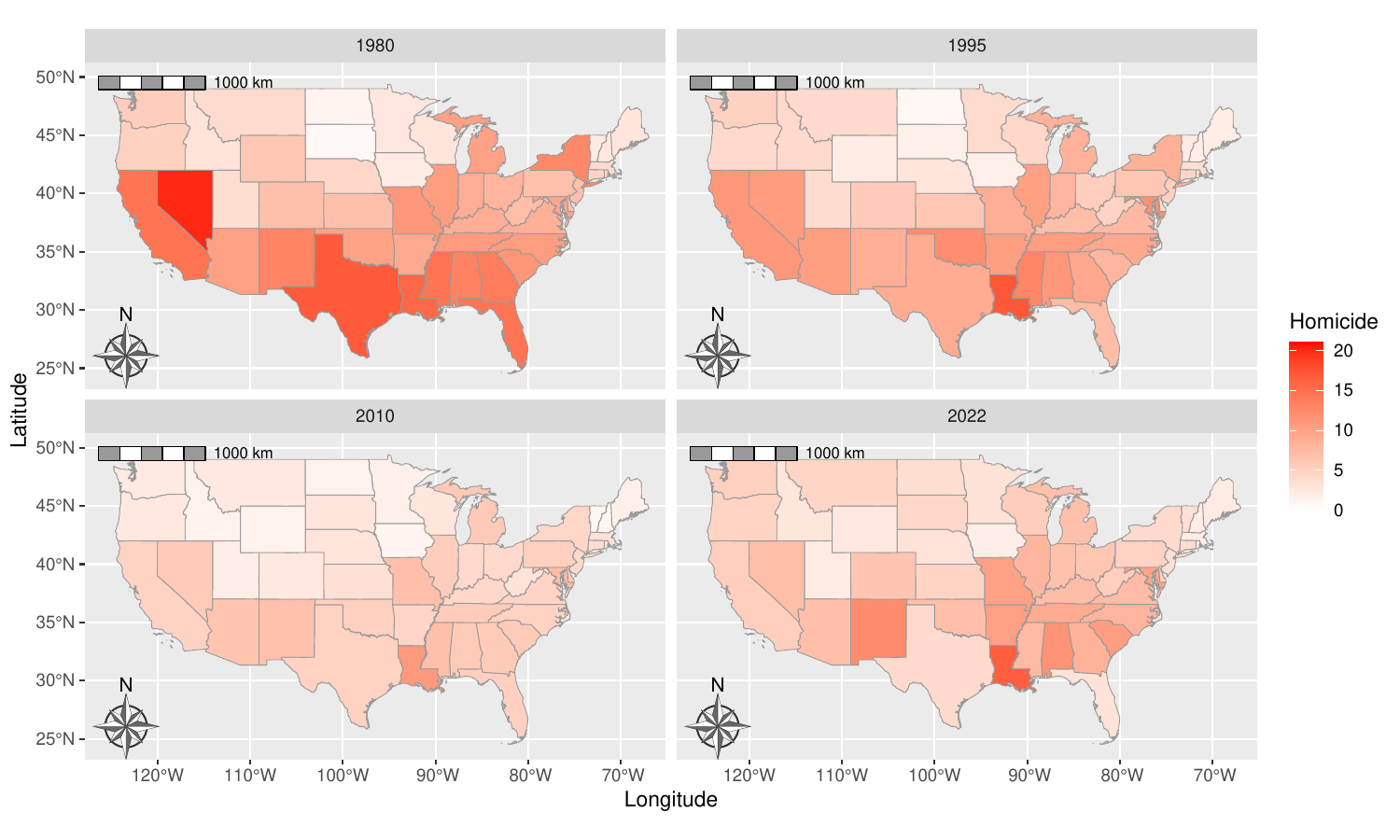}
\caption{Homicide rate per 100000 population by state in 1980, 1995, 2010 and 2022}
\label{StateHom}
\end{figure}

For this study, data on the homicide rate per 100000 population in the US states from 1979 to 2023 were collected. These data were obtained from the FBI's webpage \citep{ucr_crime_trend} annual reports\footnote{\url{https://cde.ucr.cjis.gov/LATEST/webapp/#/pages/explorer/crime/crime-trend}}. In addition, socioeconomic variables such as the percentage of people living in poverty and median income by state were included from the US Census and income and poverty tables \citep{us_census_population, us_census_poverty_data}.
Modeling crime rates around the world, particularly in the United States, is a common problem in spatial and spatio-temporal research \citep{reid2011mapping,roth2013spatiotemporal, ha2024spatial}.
Figure \ref{StateHom} shows state homicide rates for the years 1980, 1995, 2010, and 2022 in the United States. There is a decrease in the homicide rate from 1995 to 2010 and an increase in most states by 2022. Hawaii and Alaska are not shown because, as they do not have neighboring states, they cannot be included in the analysis methodology proposed in this work. 
\begin{figure}[!ht]
\centering
\includegraphics[width=14cm]{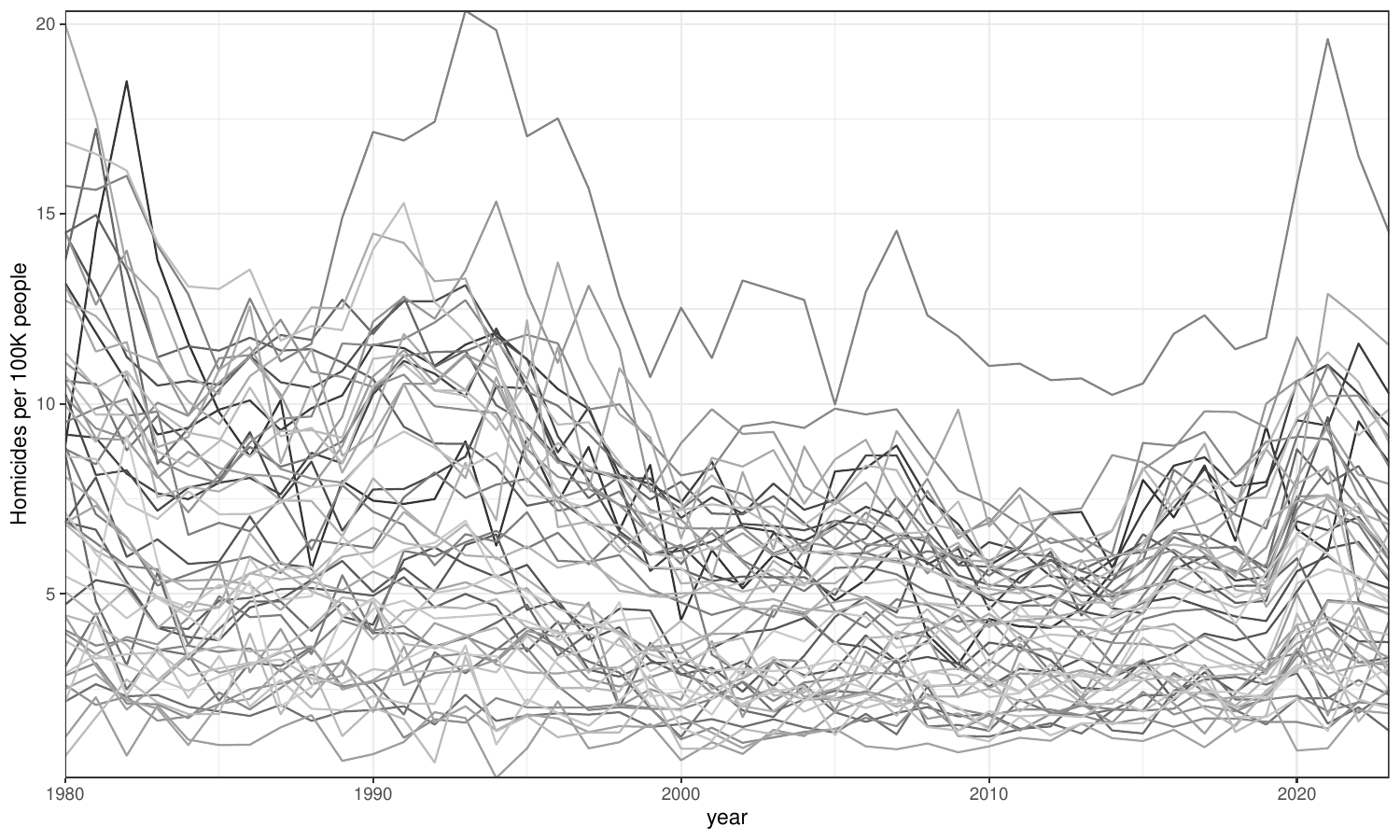}
\caption{Homicide rate per 100000 population by state across the years}
\label{StateHomLin}
\end{figure}
Figure \ref{StateHomLin} shows state homicide rates from 1980 to 2023 for each state in the United States. The same thing is evident as in Figure \ref{StateHom}. It is noteworthy that the District of Columbia was excluded because it has a very different dynamic than the other states in the US.
The proposed model is as follows:
\begin{equation}\label{USmod}
    {y}_{it}=\sum_{j=1}^{48}\rho_{j} w_{ij}y_{jt} + \sum_{k=1}^4 \beta_{k} x_{itk}  + \epsilon_{it}
\end{equation}
where $i=1, \ldots,48$ are the states of the US without Alaska and Hawaii. Since there are 44 years between 1980 and 2023 and a lag of the variable ${Y}_{it}$ is used as an explanatory variable, i.e, $x_{it4}=y_{i(t-1)}$ for capture the trend in the crime rate, as explained \citet{ha2024spatial}, then $t=1, \ldots, 43$. $x_{it1}=1$ indicates the intercept, $x_{it2}$ is the percent of people in poverty, and $x_{it2}$ is the mean income in hundred of dollars. This income is based on a 2023 baseline, as indicated by the Census office \citep{us_census_population}. 

\begin{figure}[!ht]
\centering
\includegraphics[width=12cm]{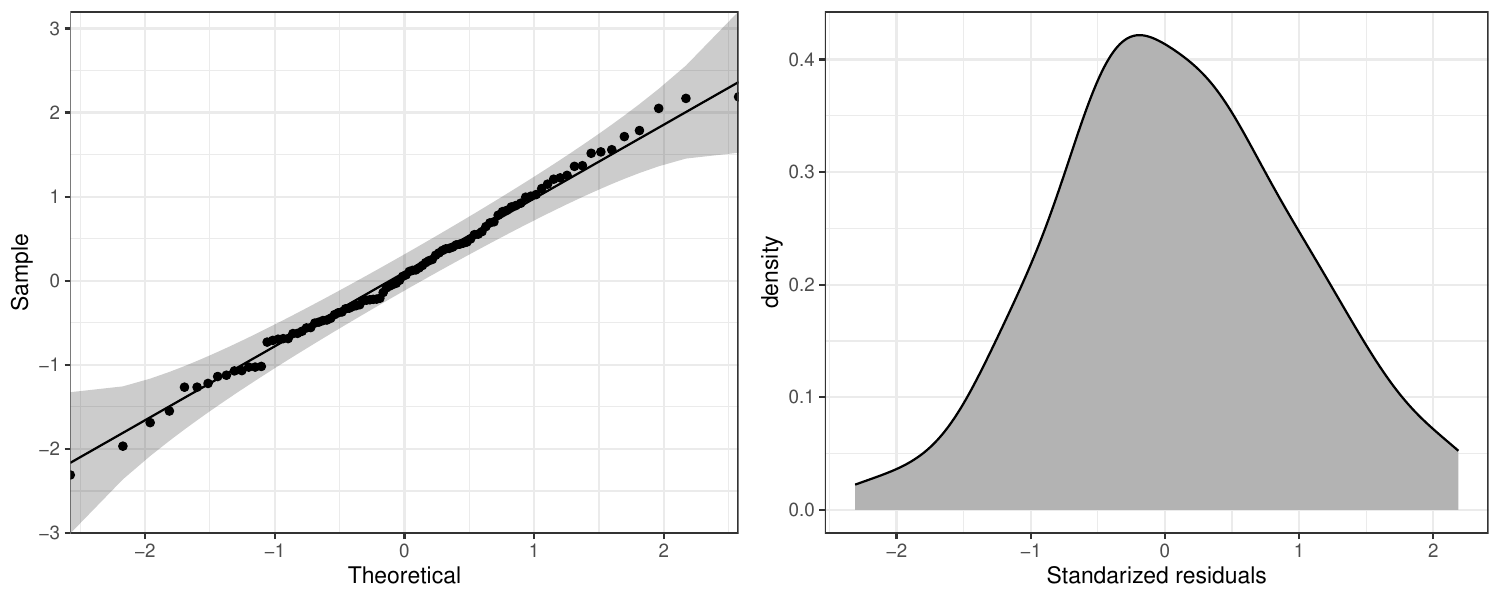}
\caption{QQplot and density for the $nT-n=2064$ standardized residuals of model estimated by Equation \eqref{USmod}}
\label{QQ}
\end{figure}

To verify that the use of specific coefficients is necessary, the $T^2$ statistic proposed in Equation \eqref{tdehotelling} is calculated. First, a usual SAR model is estimated with the \cite{piras2010sphet} library, and the following is obtained:
\begin{equation}\nonumber
\hat{\rho}=0.00457, \; \widehat{Var}(\hat{\rho})=0.0001869
\end{equation}
and with these values, it is obtained that $T^2=875324.3$ and its associated p-value is $<0.0001$; therefore, it is necessary to use specific spatial coefficients.
Before any analysis, a model proposed in section \ref{hoSPSAR} is carried out and the residuals are obtained. It is necessary to analyze the normality of the residuals, which can be seen reflected in Figure \ref{QQ}. There, it is noted that both the quantile plot and the estimated density approximate a normal distribution very well; additionally, an Anderson-Darling test ($A = 0.182$, p-value =$0.9104$) is carried out, with which it can be affirmed that there is normality.

\begin{figure}[!ht]
\centering
\includegraphics[width=14cm]{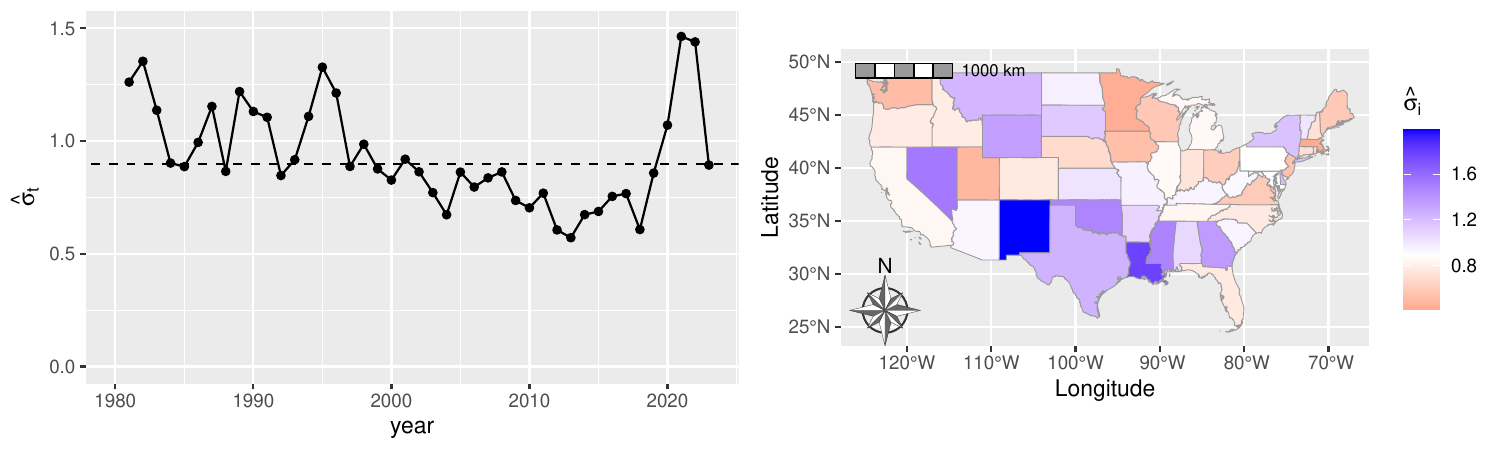}
\caption{Standard deviation of the $\hat{\epsilon}_{it}$ divided by year on left and states on right position of the model estimated by Equation \eqref{USmod}.}
\label{YearState}
\end{figure}
In Figure \ref{YearState}, the dotted lines on the years graph and the white color on the map represent the value of $\hat{\sigma}=\sqrt{0.8060}$ estimate by Equation \eqref{USmod}. 
The standard deviation across years and between states appears non-homogeneous, with values ranging from 0.7 to 1.5 for all years, and from 0.6 to 1.8 between states. Two Levene tests are conducted on the residuals using \textit{LeveneTest} of car package of \citet{Rmanual}, the first one by state ($F=6.7989$, $df=47$, $p<0.001$) and by year ($F=3.1598$, $df=42$, $p<0.001$), to evaluate the validity of the homoscedasticity assumption, which is rejected. 
\begin{table}[ht]
\centering
\begin{tabular}{|l|cccccc|}
  \hline
 Parameter& Estimate & Standard error & P-value & Direct & Indirect & Total \\ 
  \hline
 Intercept& 0.3944 & 0.1238 & $<$0.0001 &  &  &  \\ 
  Poverty & 0.0137 & 0.0073 & 0.0404 & 0.0134 & 0.0041 & 0.0175 \\ 
  Income & -0.0007 & 1.2536 & 0.9990 & -0.0007 & -0.0001 & -0.0008 \\ 
  Homicide Rate Lag & 0.7229 & 0.0078 & $<$0.0001 & 0.7342 & 0.2142 & 0.9448 \\ 
  $\sigma^2$ & 0.8860 &  0.0006&&&&\\
   \hline
\end{tabular}
\caption{Coefficients $\beta_k$ estimated for each covariate with robust estimator. The value of its standard error and the associated p-value are added and the direct, indirect, and total impacts are obtained by Equation \eqref{sk_dit}}
\label{EstiHet}
\end{table}
Therefore, the robust estimator proposed in section \eqref{hetSPSAR} is used, the estimates are shown in Table \ref{EstiHet}. However, to maintain the comparison, Table \ref{Esti} shows the estimators of the homoscedastic model, which are very similar, in line with what was obtained in the simulation.
\begin{table}[ht]
\centering
\begin{tabular}{|l|cccccc|}
  \hline
 Parameter& Estimate & Standard error & P-value & Direct & Indirect & Total \\ 
  \hline
 Intercept& 0.4544 & 0.1057 & $<$0.0001 &  &  &  \\ 
  Poverty & 0.0124 & 0.0063 & 0.0504 & 0.0125 & 0.0034 & 0.0158 \\ 
  Income & -0.0007 & 1.1236 & 0.9995 & -0.0007 & -0.0002 & -0.0009 \\ 
  Homicide Rate Lag & 0.7393 & 0.0068 & $<$0.0001 & 0.7442 & 0.2013 & 0.9454 \\ 
  $\sigma^2$ & 0.8060 &  0.0006&&&&\\
   \hline
\end{tabular}
\caption{Coefficients $\beta_k$ estimated for each covariate with homoscedastic estimator. The value of its standard error and the associated p-value are added and the direct, indirect, and total impacts are obtained by Equation \eqref{sk_dit}}
\label{Esti}
\end{table}
To analyze the effect of the covariates, it is incorrect to use only the value of the estimator as a marginal effect because, similar to what happens in the SAR model, there is spatial interaction between the different areas. Therefore, the direct, indirect (spillover), and total average effects must be calculated.
The direct effect is represented by the average of the diagonal terms of the partial derivative matrix, $\pmb{S}_k$, defined as:
\begin{equation}\label{sk_dit}
\pmb{S}_k = \beta_k(\mathbf{I}_n-\pmb{WP})^{-1}
\end{equation}
The indirect effect is the average of the off-diagonal elements in each row (or column) of the same matrix, $\pmb{S}_k$. The total effect is represented as the sum of the direct and indirect effects. These estimators can be interpreted following the ideas presented by \citet{lesage2009introduction}.

Table \ref{EstiHet} presents the estimated values for the parameters $\beta_k$, together with their direct and indirect effects estimated by Equation \eqref{sk_dit}. The estimated value of the parameter $\sigma^2$ is also presented. As this shows the adequacy of the model, the direct effects of the variables: An increase in poverty, keeping the other conditions constant, generates an increase in the homicide rate, both directly and slightly indirectly. Income does not have a significant effect, which may be due to its correlation with poverty. The homicide rate has a high positive correlation because $\hat{\beta}_4=0.7393$ with its past, which has already been explored in works by \citet{shaker2015spatial} and \citet{ha2024spatial}. 

\begin{table}[!ht]
    \centering
    \begin{tabular}{|c|c|c|}
      \hline
      \begin{tabular}{lccc}

 State & $\hat{\rho}_i$ & $se(\hat{\rho}_i)$ & P-value \\ 
  \hline
WA & -0.931 & 0.206 & $<$0.001 \\ 
  ME & -0.917 & 0.271 & 0.001 \\ 
  MD & -0.229 & 0.075 & 0.002 \\ 
  NE & -0.218 & 0.173 & 0.206 \\ 
  NM & -0.142 & 0.068 & 0.035 \\ 
  NY & -0.075 & 0.051 & 0.144 \\ 
  KS & -0.060 & 0.159 & 0.704 \\ 
  LA & -0.032 & 0.065 & 0.629 \\ 
  CT & -0.015 & 0.091 & 0.868 \\ 
  SD & -0.015 & 0.166 & 0.927 \\ 
  MI & -0.009 & 0.111 & 0.933 \\ 
  CO & 0.023 & 0.121 & 0.848 \\ 
  IL & 0.030 & 0.097 & 0.755 \\ 
  MN & 0.066 & 0.222 & 0.764 \\ 
  AL & 0.120 & 0.041 & 0.003 \\ 
  WY & 0.124 & 0.140 & 0.376 \\ 
\end{tabular} & 
      \begin{tabular}{lccc}
 State & $\hat{\rho}_i$ & $se(\hat{\rho}_i)$ & P-value \\ 
  \hline
MT & 0.142 & 0.189 & 0.453 \\ 
  SC & 0.164 & 0.109 & 0.130 \\ 
  MS & 0.165 & 0.077 & 0.033 \\ 
  MO & 0.165 & 0.065 & 0.011 \\ 
  ND & 0.173 & 0.264 & 0.513 \\ 
  ID & 0.184 & 0.143 & 0.199 \\ 
  NH & 0.197 & 0.088 & 0.024 \\ 
  IN & 0.206 & 0.119 & 0.084 \\ 
  NV & 0.207 & 0.061 & 0.001 \\ 
  VA & 0.209 & 0.110 & 0.058 \\ 
  GA & 0.209 & 0.038 & $<$0.001 \\ 
  WI & 0.214 & 0.170 & 0.208 \\ 
  AZ & 0.241 & 0.068 & $<$0.001 \\ 
  NC & 0.268 & 0.049 & $<$0.001 \\ 
  PA & 0.270 & 0.093 & 0.004 \\ 
  TN & 0.286 & 0.076 & $<$0.001 \\ 
\end{tabular}&
      \begin{tabular}{lccc}
 State & $\hat{\rho}_i$ & $se(\hat{\rho}_i)$ & P-value \\ 
  \hline
KY & 0.297 & 0.087 & 0.001 \\ 
  TX & 0.306 & 0.072 & $<$0.001 \\ 
  OR & 0.315 & 0.107 & 0.003 \\ 
  FL & 0.321 & 0.097 & 0.001 \\ 
  MA & 0.384 & 0.120 & 0.001 \\ 
  DE & 0.402 & 0.130 & 0.002 \\ 
  OK & 0.456 & 0.081 & $<$0.001 \\ 
  VT & 0.464 & 0.223 & 0.038 \\ 
  CA & 0.476 & 0.076 & $<$0.001 \\ 
  RI & 0.476 & 0.155 & 0.002 \\ 
  WV & 0.476 & 0.130 & $<$0.001 \\ 
  OH & 0.510 & 0.106 & $<$0.001 \\ 
  NJ & 0.727 & 0.157 & $<$0.001 \\ 
  AR & 0.735 & 0.074 & $<$0.001 \\ 
  IA & 0.953 & 0.164 & $<$0.001 \\ 
  UT & 0.964 & 0.070 & $<$0.001 \\ 
\end{tabular}\\
  \hline
    \end{tabular}
    \caption{Specific spatial coefficients $\rho_i$ estimated for each state with robust estimator. The value of its standard error estimated by Equation \eqref{hessrho} and the p-value associated using the result Theorem \ref{te1}}
    \label{rhosUS}
\end{table}
Table \ref{rhosUS} presents the estimates of the spatial coefficients for each state. Eleven states exhibit negative spatial coefficients, meaning they have a negative influence on the homicide rates of their neighbors, while 37 states have a positive influence. The states are ranked from the lowest value, Washington (-0.931), to the highest, Utah (0.964). Notably, 44 states have correlation values ranging from -0.25 to 0.75, which are common in spatial correlation analyses.
\begin{figure}[!ht]
\centering
\includegraphics[width=14cm]{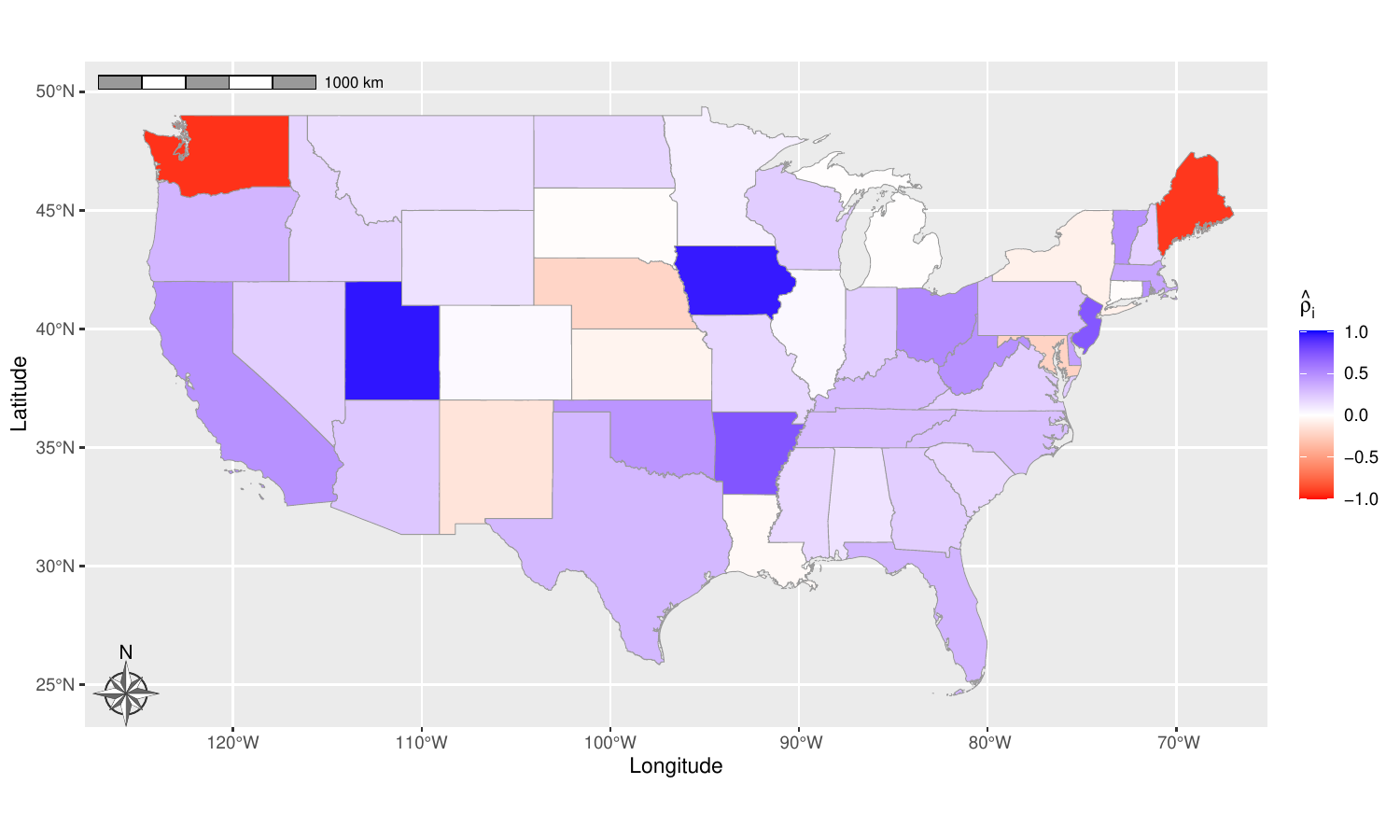}
\caption{Specific spatial coefficients $\rho_i$ estimated for each state shown in Table \ref{rhosUS}.}
\label{StateRhos}
\end{figure}
As shown in Figure \ref{rhosUS}, Washington and Maine, which exhibit very high negative correlation coefficients, are both states bordering Canada. This is likely because their American neighbors are fewer in number, with the proximity to Canada playing a significant role in shaping their spatial interactions. As a result, when the homicide rate in these states decreases, their American neighbors experience an increase in homicide rates. In contrast, Utah and Iowa, with coefficients close to 1, positively influence their neighbors' homicide rates. The remaining states show smaller positive correlations but still exhibit a general positive tendency. Crime among Mexico's neighboring states has a positive correlation, something studied in sociological phenomena of these states, such as \citet{ha2024spatial, alvarez2024crimen, slack2021postremoval}.

Notably, the only states with very high negative values are those bordering Canada, a pattern not observed in other parts of the United States. This suggests that there may be multiple factors at play, which could be further explored. The methodology used here provides a valuable starting point for analyzing crime patterns in the United States.

\section{Conclusions}
The spatial autoregressive model with specific coefficients presented in this work offers an innovative extension to standard SAR models by incorporating the spatial dependence structure differentiated by each spatial region, as it evolves over time. This approach allows capturing interactions between spatial regions in a differentiated form.
The estimation methodology for homoscedastic data and a theory for non-homoscedastic data are built.
In addition, the theoretical results guarantee consistency and efficiency in the estimators of $\pmb{\rho}$ and $\pmb{\beta}$.
The theoretical formulation is reinforced with the derivation of the asymptotic properties of the estimators, which provides a robust framework for its practical implementation. The algorithm is developed in the R software, which will allow it to become a standard tool in the analysis of complex spatial data.
The proposed model proves to be more effective than other similar techniques for modeling data with normal spatial structure and distribution, even when the residuals are not normal or homoscedastic.

For data with Gamma distribution, the proposed model also showed lower bias and higher consistency compared to other approaches, including the Ro-SARAR model. This result reinforces the robustness of the methodology for heteroscedastic distributions. In the case of the Binomial distribution, the results are consistent with the previous cases: the proposed model is unbiased and consistent, outperforming conventional methods. These results highlight the applicability of the approach to spatial data with non-normal responses, positioning it as a robust and versatile tool for spatial data analysis.

The application to modeling homicide rates in the United States shows the statistical and social analysis advantages that the model can offer.
In future work, nonlinear behaviors in the explanatory variables should be incorporated into the model using generalized additive models (GAM), and constructing a spatio-temporal heterogeneity test for data of this type.
\section*{Supplementary files}
\begin{enumerate}
    \item Supplementary file 1\label{sf1}: R files with the three simulations of this paper, the proposed methodology, and the data files.
\end{enumerate}

\appendix
\section{Gradient and Hessian of $\pmb{\rho}$}\label{apenA}
Let
\begin{align}
    \ell(\hat{\pmb{\beta}}, \pmb{\rho}, \hat{\sigma}^2)&=-\frac{nT}{2}\ln\left\{[(\mathbf{I}_{nT}-\mathbf{H})\mathbf{AY})^\top
(\mathbf{I}_{nT}-\mathbf{H})\mathbf{AY}]\right\}+\ln|\mathbf{A}| + C\nonumber
\end{align}
where $\mathbf{H}=\mathbf{X}(\mathbf{X}^{\top}\mathbf{X})^{-1}\mathbf{X}^{\top}$ and $C=-\frac{nT}{2}(1+\ln{(2\pi)})$. Note that
\begin{align}
\mathbf{A}^{-1} &= [\mathbf{I}_{nT}-(\mathbf{I}_T\otimes \mathbf{WP})]^{-1} =[\mathbf{I}_{T}\otimes(\mathbf{I}_n- \mathbf{WP})]^{-1}= \mathbf{I}_{T}\otimes(\mathbf{I}_n- \mathbf{WP})^{-1} \nonumber
\end{align}
Thus,
\begin{align}
\frac{\partial \mathbf{A}}{\partial\rho_i}&=\frac{\partial(\mathbf{I}_{nT}-(\mathbf{I}_T\otimes \mathbf{WP}))}{\partial\rho_i}=-\mathbf{I}_T\otimes \frac{\partial(\mathbf{WP})}{\partial\rho_i} = -\mathbf{I}_T\otimes \mathbf{We}_i\mathbf{e}_i^\top\nonumber
\end{align}
where $\mathbf{e}_i = (0,\ldots, 0, \overset{i}{1}, 0, \ldots, 0)$ is the $i$-th canonical vector of $\mathbb{R}^n$. Then,
\begin{align}
  \frac{\partial\ln \vert \mathbf{A}\vert}{\partial\rho_i} &= \operatorname{tr}\left(\mathbf{A}^{-1}\frac{\partial\mathbf{A}}{\partial\rho_i}\right)=-\operatorname{tr}\left\{\mathbf{A}^{-1}[\mathbf{I}_T\otimes \mathbf{We}_i\mathbf{e}_i^\top]\right\}\nonumber \\
  &=-\operatorname{tr}\left\{[\mathbf{I}_{T}\otimes(\mathbf{I}_n- \mathbf{WP})^{-1} )(\mathbf{I}_T\otimes \mathbf{We}_i\mathbf{e}_i^\top]\right\} \nonumber \\
  &=-\operatorname{tr}\left((\mathbf{I}_{T}\otimes(\mathbf{I}_n- \mathbf{WP})^{-1}  \mathbf{We}_i\mathbf{e}_i^\top)\right)\nonumber \\
  &=-T\left[(\mathbf{I}_n- \mathbf{WP})^{-1}  \mathbf{W}\right]_{ii} \label{part1drho}
\end{align}
The second partial derivative with respect to $\rho_j$ is:
\begin{align}
      \frac{\partial^2\ln \vert \mathbf{A}\vert}{\partial\rho_j\partial\rho_i}&=\frac{\partial}{\partial\rho_j}\left(-T\left[(\mathbf{I}_n- \mathbf{WP})^{-1}  \mathbf{W}\right]_{ii}\right)\nonumber\\
      &=T\left[(\mathbf{I}_n- \mathbf{WP})^{-1}\frac{\partial\mathbf{WP}}{\partial\rho_j}(\mathbf{I}_n- \mathbf{WP})^{-1}\mathbf{W}\right]_{ii}\nonumber\\
      &=T\left[(\mathbf{I}_n- \mathbf{WP})^{-1}\mathbf{We}_j\mathbf{e}_j^\top(\mathbf{I}_n- \mathbf{WP})^{-1}\mathbf{W}\right]_{ii}\nonumber\\
      &=T\left[(\mathbf{I}_n- \mathbf{WP})^{-1}\mathbf{We}_j\mathbf{e}_j^\top\right]_{ij}\left[(\mathbf{I}_n- \mathbf{WP})^{-1}\mathbf{W}\right]_{ji}\label{part1ddrho}
\end{align}
Thus, it is obtained that:
\begin{align}
      \frac{\partial^2\ln \vert \mathbf{A}\vert}{\partial\pmb{\rho}^\top\partial\pmb{\rho}}&=T \operatorname{diag}\left[(\mathbf{I}_n- \mathbf{WP})^{-1}\mathbf{We}_j\mathbf{e}_j^\top\right]\operatorname{diag}\left[(\mathbf{I}_n- \mathbf{WP})^{-1}\mathbf{We}_j\mathbf{e}_j^\top\right]^\top\nonumber
\end{align}
Let $\mathbf{M}=\mathbf{AY}$, it is clear that:
\begin{align}
    \frac{\partial\mathbf{M}}{\partial\rho_i} &= \frac{\partial\mathbf{A}}{\partial\rho_i}\mathbf{Y}=-(\mathbf{I}_T\otimes \mathbf{We}_i\mathbf{e}_i^\top)\mathbf{Y}\nonumber
\end{align}
Therefore, 
\begin{align}
     &\frac{\partial\ln(\mathbf{M}^\top(\mathbf{I}_{nT}-\mathbf{H})\mathbf{M} )}{\partial\rho_i}=\left[\mathbf{M}^\top(\mathbf{I}_{nT}-\mathbf{H})\mathbf{M} \right]^{-1}\frac{\partial}{\partial\rho_i}\operatorname{tr}\left((\mathbf{I}_{nT}-\mathbf{H})\mathbf{M}\mathbf{M}^\top\right)\nonumber\\
     &=-\left[\mathbf{M}^\top(\mathbf{I}_{nT}-\mathbf{H})\mathbf{M} \right]^{-1}\operatorname{tr}\Big[(\mathbf{I}_{nT}-\mathbf{H})(\mathbf{I}_T\otimes \mathbf{We}_i\mathbf{e}_i^\top)\mathbf{Y}\mathbf{M}^\top+\nonumber\\
   & (\mathbf{I}_{nT}-\mathbf{H})\mathbf{M}\mathbf{Y}^\top\mathbf{e}_i\mathbf{e}_i^\top \mathbf{W}^\top \Big]\label{part2drho}
\end{align}
The second derivative is obtained as:
\begin{align}
     &\frac{\partial^2\ln(\mathbf{M}^\top(\mathbf{I}_{nT}-H)\mathbf{M} )}{\partial\rho_j\partial\rho_i}=\left[\mathbf{M}^\top(\mathbf{I}_{nT}-\mathbf{H})\mathbf{M} \right]^{-2}\frac{\partial}{\partial\rho_j}\operatorname{tr}\left[(\mathbf{I}_{nT}-\mathbf{H})\mathbf{M}\mathbf{M}^\top\right]\times \nonumber\\
     &\operatorname{tr}\Big[ (\mathbf{I}_{nT}-\mathbf{H})(\mathbf{I}_T\otimes \mathbf{We}_i\mathbf{e}_i^\top)\mathbf{Y}\mathbf{M}^\top+(\mathbf{I}_{nT}-\mathbf{H})\mathbf{M}\mathbf{Y}^\top\mathbf{e}_i\mathbf{e}_i^\top \mathbf{W}^\top \Big] +\nonumber\\
     &\left[\mathbf{M}^\top(\mathbf{I}_{nT}-\mathbf{H})\mathbf{M} \right]^{-1} \frac{\partial}{\partial\rho_j}\operatorname{tr}\Big[ (\mathbf{I}_{nT}-\mathbf{H})(\mathbf{I}_T\otimes \mathbf{We}_i\mathbf{e}_i^\top)\mathbf{Y}\mathbf{M}^\top+\nonumber\\
     &(\mathbf{I}_{nT}-\mathbf{H})\mathbf{M}\mathbf{Y}^\top\mathbf{e}_i\mathbf{e}_i^\top \mathbf{W}^\top \Big]\label{part2ddrho}
\end{align}
Combining equations \eqref{part1drho} and \eqref{part2drho}, Equation \eqref{drho}  is obtained. Also, combining equations \eqref{part1ddrho} and \eqref{part2ddrho}, Equation \eqref{hessrho} is obtained.
\section{Approximation to information matrix for $\pmb{\rho}$}\label{apenB}
Let $f(x,y):\mathbb{R}^2\to \mathbb{R}$ continouos and derivable in a open set $\mathcal{A}\in \mathbb{R}^2$ then it is possible aproximate $f$ near $(x_0, y_0)\in \mathcal{A}$ by the next expression \citep{jiang2010large}:
\begin{align}
    f(x,y)&\approx f(x_0, y_0)+\frac{\partial f}{\partial x}\left(x_0, y_0\right)(x-x_0)+\frac{\partial f}{\partial y}\left(x_0, y_0\right)(y-y_0)+\nonumber\\
    &\frac{1}{2}\Big[\frac{\partial^2 f}{\partial x^2}\left(x_0, y_0\right)(x-x_0)^2+\frac{\partial^2 f}{\partial y^2}\left(x_0, y_0\right)(y-y_0)^2+\nonumber\\
    & 2\frac{\partial^2 f}{\partial x\partial y}\left(x_0, y_0\right)(x-x_0)(y-y_0)\Big]\label{taylor}
\end{align}
Therefore, using Equation \eqref{taylor}, it is obtained that:
\begin{align}
    \mathbb{E}\left(\frac{X}{Y}\right) \approx \frac{\mathbb{E}(X)}{\mathbb{E}(Y)}-\frac{\operatorname{Cov}(X,Y)}{\mathbb{E}(Y)^2}+\frac{\operatorname{Var}(Y)\mathbb{E}(X)}{\mathbb{E}(Y)^3}\label{taylor1}
\end{align}
\begin{equation}
    \mathbb{E}\left(\frac{X}{Y^2}\right) \approx \frac{\mathbb{E}(X)}{\mathbb{E}(Y)^2}+\frac{3\operatorname{Var}(Y)\mathbb{E}(X)}{\mathbb{E}(Y)^4}-\frac{2\operatorname{Cov}(X,Y)}{\mathbb{E}(Y)^3}+\frac{3\operatorname{Var}(Y)\mathbb{E}(X)}{\mathbb{E}(Y)^4}\label{taylor3}
\end{equation}
Using the fact $\mathbf{AY}\sim \mathbf{N}_{nT}(\mathbf{X}\pmb{\beta}, \sigma^2\mathbf{I}_{nT})$ the next equations are true:
\begin{align}
    \mathbf{E}[\mathbf{Y}^\top\mathbf{A}^\top (\mathbf{I}_{nT}-H)\mathbf{AY}]&=\operatorname{tr}\left((\mathbf{I}_{nT}-\mathbf{H})\mathbb{E}(\mathbf{AY}\mathbf{Y}^\top\mathbf{A}^\top)\right)\nonumber\\
    &=\operatorname{tr}\left[(\mathbf{I}_{nT}-\mathbf{H})[\sigma^2\mathbf{I}_{nT}+\mathbf{X}\pmb{\beta}\pmb{\beta}^\top \mathbf{X}^\top]\right]\nonumber\\
    \mathbf{E}(\mathbf{Y}\mathbf{Y}^\top)&=\mathbf{A}^{-1}\mathbf{X}\pmb{\beta}\pmb{\beta}^\top \mathbf{X}^\top\mathbf{A}^{-\top} + \sigma^2 \mathbf{A}^{-1}\mathbf{A}^{-\top}
\end{align}
with equations \eqref{taylor1}, \eqref{taylor3} and the properties of expectation, Equation \eqref{hessianrho} can be obtained.
\section{Proof of Theorem \ref{the2}}\label{apenC}
\begin{proof}
In the model proposed in Equation \eqref{PSAR} and using Equation \eqref{kj20}, it is obtained that:
\begin{equation*}
    \mathbf{Y} = \mathbf{Z}\pmb{\delta} + \pmb{\epsilon}
\end{equation*}
and hence:
\begin{align*}
    (nT)^{-\frac{1}{2}}(\hat{\pmb{\delta}}-\pmb{\delta})&= (nT)^{-\frac{1}{2}}(\left[(\mathbf{P_H}\mathbf{Z})^\top\mathbf{Z}\right]^{-1}(\mathbf{P_H}\mathbf{Z})^\top\mathbf{Y}-\pmb{\delta})\\
    &=(nT)^{-\frac{1}{2}}\left(\left[(\mathbf{P_H}\mathbf{Z})^\top\mathbf{Z}\right]^{-1}(\mathbf{P_H}\mathbf{Z})^\top\mathbf{Z}(\pmb{\delta} + \pmb{\epsilon})-\pmb{\delta}\right)\\
    &=(nT)^{-\frac{1}{2}}\left[(\mathbf{P_H}\mathbf{Z})^\top\mathbf{Z}\right]^{-1}(\mathbf{P_H}\mathbf{Z})^\top\mathbf{Z}\pmb{\epsilon}
\end{align*}
Using assumptions (6) and (7), it follows that:
\begin{align*}
    (nT)^{-1}(\mathbf{P_H}\mathbf{Z})^\top\mathbf{Z}-\mathbf{Q_{HZ}}^\top \mathbf{Q_{HH}}^{-1}\mathbf{Q_{HZ}}=o_{n+K}(1) \text{ as } T\to \infty
\end{align*}
Since the assumption (6) and (7), it is obtained that: 
\begin{align*}
    \mathbf{Q_{HZ}}^\top \mathbf{Q_{HH}}^{-1}\mathbf{Q_{HZ}}&=\mathcal{O}_{n+K}(1)\text{ as } T\to \infty\\
    \left[\mathbf{Q_{HZ}}^\top \mathbf{Q_{HH}}^{-1}\mathbf{Q_{HZ}}\right]^{-1}&=\mathcal{O}_{n+K}(1) \text{ as } T\to \infty
\end{align*}
Thus,
\begin{align*}
    \left[(nT)^{-1}\mathbf{P_H}\mathbf{Z})^\top\mathbf{Z}\right]^{-1}-\left[\mathbf{Q_{HZ}}^\top \mathbf{Q_{HH}}^{-1}\mathbf{Q_{HZ}}\right]^{-1}&=o_{n+K}(1) \text{ as } T\to \infty
\end{align*}
Furthermore, it follows that:
\begin{align*}
    &\mathbf{Q_{HH}}^{-1}\mathbf{Q_{HZ}}^\top[\mathbf{\mathbf{Q_{HZ}}^\top Q_{HH}}^{-1}\mathbf{Q_{HZ}}]^{-1} = o_{n+K}(1) \text{ as } T\to \infty
\end{align*}
and it follows that:
\begin{align*}
    &(n^{-1}\mathbf{H}^\top\mathbf{H})^{-1}(n^{-1}\mathbf{H}^\top\mathbf{Z})(n^{-1}\mathbf{Z}^\top\mathbf{H})(n^{-1}\mathbf{H}^\top\mathbf{H})^{-1}(n^{-1}\mathbf{H}^\top\mathbf{Z})^{-1}-\\
    &\mathbf{Q_{HH}}^{-1}\mathbf{Q_{HZ}}^\top[\mathbf{\mathbf{Q_{HZ}}^\top Q_{HH}}^{-1}\mathbf{Q_{HZ}}]^{-1} = o_{n+K}(1)\text{ as } T\to \infty
\end{align*}
Now, from Equation \eqref{PSAR}, and assumptions (1), (2), and (3) it is obtained that $\sup\vert \vert\pmb{\beta}\vert\vert<\infty$.

Therefore, all columns of $\mathbf{Z}$ are of the form $\pmb{\gamma}+\pmb{\Gamma}\pmb{\epsilon}$,  where $\pmb{\gamma}$ is a vector of size $nT\times 1$ whose elements are bounded in absolute value, and the row and column sums of the matrix $\pmb{\Gamma}$ of size $nT\times nT$  are uniformly bounded in absolute value by some finite constant.
From assumption (5), $\mathbf{H}$ is uniformly bounded in absolute value. By assumption (4), $\mathbb{E}(\pmb{\epsilon})=\mathbf{0}$ and its diagonal variance matrix $\pmb{\Sigma}$ are uniformly bounded elements.
Therefore, $\mathbb{E}((nT)^{-\frac{1}{2}}\mathbf{H}^\top\pmb{\epsilon})=\mathbf{0}$ and the elements of 
\begin{align*}
    Var((nT)^{-\frac{1}{2}}\mathbf{H}^\top\pmb{\epsilon})=(nT)^{-1}\mathbf{H}^\top\pmb{\Sigma}\mathbf{H}
\end{align*}
are also uniformly bounded in absolute value. Let $\pmb{\nu}=(nT)^{-\frac{1}{2}}\mathbf{H}^\top\pmb{\epsilon}$, then $\mathrm{E}(\pmb{\nu})=\mathbf{0}$ and $\mathrm{Var}(\pmb{\nu})=\mathbf{V}$, with $\mathbf{V}$ a positive definite matrix. Thus, by multivariate Chebyshevs inequality \citep{Chebis}, it is clear that $\forall v>0$:
\begin{align*}
    P\left(\sqrt{\pmb{\nu}^\top \mathbf{V}^{-1}\pmb{\nu}}>v \right)\leq \frac{n+K}{Tv^2}
\end{align*}
Hence, $(nT)^{-\frac{1}{2}}\mathbf{H}^\top\pmb{\epsilon}=\mathcal{O}_{n+K}(1) \text{ as } T\to \infty$, and consequently:
\begin{align*}
    (nT)^{-\frac{1}{2}}(\hat{\pmb{\delta}}-\pmb{\delta})=\mathbf{B}\pmb{\epsilon}+&o_{n+K}(1)\text{ as } T\to \infty
\end{align*}
where $\mathbf{B}=\mathbf{Q_{HH}}^{-1}\mathbf{Q_{HZ}}^\top[\mathbf{\mathbf{Q_{HZ}}^\top Q_{HH}}^{-1}\mathbf{Q_{HZ}}]^{-1}\mathbf{H}^\top$. By the assumption (1) and (4), the conditions for Theorem 1 of \cite{rothenberg1984approximate} are satisfied, and hence:
\begin{equation*}
     \hat{\pmb{\delta}}\xrightarrow[T\to \infty]{d} N_{K+n}\left(\pmb{\delta}, \frac{1}{T}\mathbf{B}^\top\mathbf{B}\right)
\end{equation*}
\end{proof}
\end{document}